\documentclass[lettersize,journal]{IEEEtran}
\usepackage{amsmath,amsfonts}
\usepackage{array}
\usepackage{textcomp}
\usepackage{stfloats}
\usepackage{url}
\usepackage{verbatim}
\usepackage{graphicx}
\usepackage{cite}

\usepackage{color,url}
\usepackage{algorithm,algpseudocode}
\usepackage{amssymb,mathtools}
\usepackage{caption}
\usepackage{multirow}
\usepackage{bm,bbm}
\usepackage{booktabs}
\usepackage[caption=false,font=footnotesize,labelformat=empty]{subfig}
\usepackage[whole]{bxcjkjatype}
\graphicspath{{./_figs/}}

\newcommand{\argmin}{\mathop{\mathrm{arg~min}}\limits}

\algnewcommand\algorithmicinput{\textbf{Input:}}
\algnewcommand\Input{\item[\algorithmicinput]}

\algnewcommand\algorithmicmyreturn{\textbf{Return:}}
\algnewcommand\Myreturn{\item[\algorithmicmyreturn]}

\algnewcommand\algorithmicdef{\textbf{Define}}
\algnewcommand\Mydef{\item[\algorithmicdef]}

\newcommand{\relmiddle}[1]{\mathrel{}\middle#1\mathrel{}}

\begin{document}

\title{Computer Vision-assisted Single-antenna and Single-anchor RSSI Localization Harnessing Dynamic Blockage Events}

\author{Tomoya~Sunami, \IEEEmembership{Student~Member,~IEEE},
Sohei~Itahara, \IEEEmembership{Student~Member,~IEEE},
Yusuke~Koda, \IEEEmembership{Graduate~Student~Member,~IEEE},
Takayuki~Nishio, \IEEEmembership{Senior~Member,~IEEE},
  and Koji~Yamamoto, \IEEEmembership{Senior~Member,~IEEE}.
\thanks{This work was supported in part by MIC/SCOPE \#JP196000002.}
\thanks{T.~Sunami, S.~Itahara, and K.~Yamamoto are with Graduate School of Informatics, Kyoto University, Kyoto 606-8501, Japan. (e-mail: kyamamot@i.kyoto-u.ac.jp). }
\thanks{Y.~Koda is with Centre of Wireless Communications, University of Oulu, 90014 Oulu, Finland. (e-mail: Yusuke.Koda@oulu.fi)}
\thanks{T.~Nishio is with School of Engineering, Tokyo Institute of Technology Ookayama, Meguro-ku, Tokyo, 152-8550, Japan. (e-mail: nishio@ict.e.titech.ac.jp)}}

\markboth{Journal of \LaTeX\ Class Files,~Vol.~14, No.~8, August~2021}%
{Shell \MakeLowercase{\textit{et al.}}: A Sample Article Using IEEEtran.cls for IEEE Journals}

\IEEEpubid{0000--0000/00\$00.00~\copyright~2021 IEEE}

\maketitle

\begin{abstract}
  This paper demonstrates the feasibility of single-antenna and single-RF (radio frequency)-anchor received power strength indicator (RSSI) localization (SARR-LOC) with the assistance of the computer vision (CV) technique.
	Generally, to perform radio frequency (RF)-based device localization, either 1) fine-grained channel state information or 2) RSSIs from multiple antenna elements or multiple RF anchors (e.g., access points) is required.
	Meanwhile, owing to deficiency of single-antenna and single-anchor RSSI, which only indicates a coarse-grained distance information between a receiver and a transmitter, realizing localization with \textit{single-antenna and single-anchor RSSI} is challenging.
	Our key idea to address this challenge is to leverage CV technique and to estimate the most likely first Fresnel zone (FFZ) between the receiver and transmitter, where the role of the RSSI is to detect blockage timings.
	Specifically, historical positions of an obstacle that dynamically blocks the FFZ are detected by the CV technique, and we estimate positions at which a blockage starts and ends via a time series of RSSI. 
	These estimated obstacle positions, in principle, coincide with points on the FFZ boundaries, enabling the estimation of the FFZ and localization of the transmitter.
	The experimental evaluation revealed that the proposed SARR-LOC achieved the localization error less than 1.0\,m in an indoor environment, which is comparable to that of a conventional triangulation-based RSSI localization with multiple RF anchors.
\end{abstract}


\section{Introduction}
\label{sec:introduction}
\IEEEPARstart{R}{adio} frequency (RF) localization has attracted a lot of attention.
RF localization will greatly support location-aware services in a scenario where signals in global positioning systems are not accessible, which is typified by indoor environments.
Indoor location-aware services, such as indoor navigation and surveillance, will enhance our daily lives in a wide variety of point-of-views exemplified by health, safety, and utility.
Hence, it is important to design successful RF localization systems as regards accuracy, hardware costs, and implementation ability.

Traditionally, RF localization is designed using received signal strength indicator (RSSI) from a satisfactory number of RF anchors (i.e., access points (APs)).
The reason for this requirement of multiple RF anchors is that the RSSI from a single RF anchor generally does not determine the angle of departure (AoD), which necessitates the use of multiple RF anchors.
For example, RSSI from a single RF anchor gives information on the distance between the RF anchor and transmitter only; hence, to determine the location of the transmitter, at least three RF anchors are required~\cite{gu2009survey}.
Currently, some works~\cite{passafiume2017enhanced,li2019cost} proposed RSSI localization with a single RF anchor equipped with multiple antenna elements,
by using each antenna element as independent RF anchor.
Meanwhile, targeted transmitters are not necessarily surrounded by such redundant numbers of RF anchors or an RF anchor with redundant numbers of antenna elements.
Consequently, our interest is in designing a localization method that can be built on a single RF anchor equipped with a single antenna element.

Importantly, with channel state information (CSI), a single RF anchor and even a single antenna element can be used for the localization.
The effective method of localization based on a single AP's CSI is to leverage multiple antennas and hence, calculate the AoD from antenna separations and phase differences of received signals~\cite{wielandt2017indoor}.
To estimate the AoD with higher granularity, more sophisticated algorithms typified by the multiple signal classification (MUSIC) algorithm~\cite{schmidt1986multiple} can be leveraged when the RF anchor implements many antenna elements~\cite{sen2013avoiding}.
Conversely, in the case of localization with a single antenna element, the basic idea is to leverage the channel decorrelations and hence, detect multipath characteristics from frequency-domain decorrelations or Doppler shifts from time-domain decorrelations.
For an example of a former case, the work in~\cite{grosswindhager2018salma} performed a localization exploiting reflected paths estimated by the channel impulse response and a predefined floor plan.
To leverage the time-domain channel decorrelations, the works in~\cite{komamiya2019single, komamiya2020radiation} determined the AoD with the phase shift and the displacement of the receiver determined by a Doppler shift.

As discussed above, to perform RF-based localization, at least one of the following information should be available:
1) RSSIs from multiple antenna elements or multiple RF anchors and
2) CSI from a single antenna element that can detect channel decorrelations.
However, the success of employing such strategies depends on wireless systems or channel environments.
For example, employing many antenna elements naturally requires transmitters and receivers to implement such antenna elements, which do not apply to non-high-throughput wireless communications typified by wireless nodes in sensor networks.
The second strategy for detecting channel decorrelations depends on the channel bandwidth or symbol rate.
More specifically, the strategy is available only when the following condition is roughly satisfied: $B\geq 1/\tau$ or $T \geq 1/v$, where $B$ and $T$ are the bandwidth and symbol period, respectively, and $\tau$ and $v$ denote the delay spread and the receiver speed, respectively.
Hence, this is not applicable to a scenario where the bandwidth and symbol period are smaller than the delay spread or Doppler shift, respectively, meaning the second strategy cannot necessarily be deployed on every wireless system.

These issues motivated us to design an RF localization, which is agnostic to wireless systems on which localization methods are built.
Based on the above discussion, this objective first necessitates the design of an RF localization method using a single RF anchor with a single antenna element.
Moreover, the best way to achieve this goal is to revisit the RSSI. 
Because RSSI information can be obtained in most wireless systems,
whereas fine-grained CSI that can detect channel decorrelations is not always available, which depends on the protocol, RF chipsets, the bandwidth, and the symbol rate.

Instead of the benefits of agnostics of single-antenna and single-anchor RSSI based localization, as far as we know, there is any localization method that using only single-antenna and single-anchor RSSI.  
This is because, in principle, the single-antenna and single-anchor RSSI only informs a coarse-grained distance between the RF anchor and transmitter, and AoA or AoD cannot be estimated.
Therefore, the single-antenna and single-anchor RSSI is deficient to conduct localization.
Hence, the key question in this work is; \textit{Is it feasible to perform single-antenna and single-anchor RSSI localization with the assistance of other modalities?}

Our key idea to address this challenge is to leverage a computer vision (CV) technique and to estimate the most likely first Fresnel zone (FFZ) between 
a target transmitter device and an RF anchor, which is referred to as a geometric area where the signal strength sharply decreases when an obstacle exists therein~\cite{zhang2018from}.
We termed this localization method the single-antenna and single-RF-anchor RSSI localization (SARR-LOC).
In SARR-LOC, we use RSSI from a single antenna only to the timing of the start and end of the FFZ blockage between the RF anchor and the target device caused by moving obstacles (e.g., pedestrians), which degrades the RSSI, and we do not estimate geometry-related information, such as distances and AoDs, directly from RSSI.
The geometry-related information is detected by the CV technique as the position of the obstacle that blocks the FFZ between the RF anchor and the target device.
By leveraging the obstacle positions and blockage timings estimated from RSSI,
we record blockage points, which are referred to as the position of the obstacle when the FFZ blockage occurs between the RF anchor and the target device.
These historical blockage points, in principle, coincide with the FFZ boundary, and the location of the target device can be determined by estimating the FFZ.

Note that our idea is applicable to scenarios of incomplete camera images, in which the images do not reflect on the transmitter location.
This is because our localization methodology only requires camera images to reflect on the positions of blockage obstacles, not those of target transmitters; hence, our methodology is applicable to the scenario where the target transmitter is a tiny module such as RFID chips or is packed with some boxes.
This characteristic is different from other CV-based localization techniques, which often require images to reflect on the positions of target objects.
In a nutshell, in our localization method, RF and visual information are leveraged in a highly complementary manner, and to the best of our knowledge, such an interactive perspective between these two different types of information for transmitter localizations has not been provided in the literature.

\subsection{Our Contributions}
\label{ssec:contribution}
The salient contributions are summarized as follows:
\begin{itemize}
	\item With the assistance of the CV technique, the feasibility of RF-based device localization with \textit{deficient} RF information is demonstrated.
	      For the demonstration, we performed testbed experiments using a pair of wireless local area network (WLAN) transmitter/receivers\footnote{In the experimental evaluation of this paper, the received power of the WLAN device with multiple antennas is used as the received power of the receiver with a single antenna.
		  The measured received power is affected by multi-antennas effects (e.g., diversity reception) that do not occur in case of a single antenna. 
		  However, we do not use these multi-antenna effects, explicitly.
		  So, the experiments are sufficient to show the feasibility of the proposed single antenna localization.} and an RGB-D camera with YOLO CV software and showed localization error less than 1.0\,m with a time series of RSSI from a single RF anchor only (\textbf{Fig.~\ref{fig:main_results}}).
	      Note that our examined RF localization system is highly agnostic to underlying wireless systems (e.g., number of antenna elements of the receiver and transmitter, channel bandwidth, and symbol rate) and does not require fine-grained RF information.
	      We believe that this can be built upon a wide variety of wireless systems.
	      To the best of our knowledge, this is the first study to design and experiment an RF localization possessing the aforementioned characteristics.
	\item We verified that from time series of RSSI, one can estimate whether the obstacle overlaps to the first Fresnel zone (FFZ) or not, leveraging the principle that the received power is sharply decreased, when the obstacle overlaps to the FFZ.
	      In the experiment evaluation, an empirical method estimate the overlaps of the obstacle and the FFZ with 90\% and higher accuracy (\textbf{Table~\ref{tab:miss_detection}, Fig~\ref{fig:blockage_detection_result}, and Fig.~\ref{fig:fresnel_point_estimation}~(a)}).
	\item We develop a feasible localization method with a single-antenna and single-anchor RSSI.
	      The key idea is to harness the dynamic blockage events that occur in a link between an RF anchor and the target device.
	      More specifically, we estimate the FFZ boundaries form the obstacle positions when blockage starts or ends, which is obtained from time series of RSSI.
	      Hence, the target device is localized as the edge of the elliptical shaft of the FFZ.
	      We believe that these methodological viewpoints and in-depth discussions on localization accuracy have not been provided  in the literature.  
\end{itemize}

The main scope of this study is to demonstrate the feasibility of an RF localization method with a single-antenna and single-anchor RSSI assisted by a CV technique in an example environment, where the drastic blockage by the moving obstacle effects on the RSSI.
Consequently, we examined our designed localization method in an ideal environment, wherein the RF anchor and target device were fixed until the localization procedure was completed and a large obstacle passes between the RF anchor and the target device.
This experiment suffices to achieve the goal of demonstrating feasibility of the proposed method, and we believe that the insights from this experiment are satisfactory for validating the above contribution.
Hence, experiments for other environments are beyond the scope of this study.

Moreover, it should be noted that, to the best of our knowledge,
any RF localization method using only single-antenna and single-anchor RSSI has not been proposed due to its deficiency of the single-antenna RSSI in terms of the localization.
More concretely, single-antenna RSSI only informs a distance between the RF anchor and transmitter, and it does not inform AoA or AoD, which results in the difficulty in the localization. 
Thus, we focus on the demonstration of the feasibility of an RF localization method with a single-antenna RSSI,
and avoid comparing the localization accuracy of the proposed method with existing localization methods which require CSI or RSSI from multiple anchors.

\section{Related Works}

\begin{table*}[t!]
	\caption{
		RSSI-based localization. FP indicates a RSSI fingerprinting.
		In-room and room-level indicate the less than 10\,m and more than 10\,m localization errors, respectively.
	}
	\label{table:related_works_RSSI}
	\centering
	\scalebox{1.}{
		\begin{tabular}{cccccccc}
			\toprule
			Authors                                                 & Using FP    & Single       & Single       & Single       & Accuracy                    \\
			                                                        &             & RF anchor    & Antenna      & Channel      &                           & \\
			\midrule
			Paramvir~\textit{et al.}\cite{bahl2000radar}            & No          & No           & Yes          & Yes          & In-room              \\
			Passafiume~\textit{et al.}\cite{passafiume2017enhanced} & No          & Yes          & No           & No           & In-room            \\
			Li~\textit{et al.}\cite{li2019cost}                     & Yes         & Yes          & No           & Yes          & In-room             \\
			Bai~\textit{et al.}\cite{bai2014new}                    & Yes         & No           & Yes          & Yes          & Room-level   \\
			\textbf{Proposed method}                                & \textbf{No} & \textbf{Yes} & \textbf{Yes} & \textbf{Yes} & \textbf{In-room}          \\
			\bottomrule
		\end{tabular}
	}
\end{table*}

\begin{table*}[t!]
	\caption{
		Single RF anchor with single-antenna localization systems and methods.
	}
	\label{table:related_works_SAL}
	\centering
	\scalebox{1.}{
		\begin{tabular}{cccccccc}
			\toprule
			System/method                                                      & RF Data           & Accuracy                  & Features                    \\
			\midrule
			Synthetic aperture\cite{komamiya2019single, komamiya2020radiation} & Doppler sift, AoA & In-room             & The AP is moving            \\
			Bernhard~\textit{et al.}\cite{grosswindhager2018salma}             & CIR               & In-room                 & Uses a floor plan           \\
			\textbf{Proposed method}                                           & \textbf{RSSI}     & \textbf{In-room}        & \textbf{Uses camera image}  \\
			\bottomrule
		\end{tabular}
	}
\end{table*}

We provide a brief review of related studies, detailing the difference from such studies.
First, it should be noted that from this review, we intentionally exclude the studies on \textit{fingerprinting} methods~\cite{yang2013rssi}, wherein an RF transmitter is localized via pattern matching to a database formed with a pre-obtained training dataset.
This is because our proposed localization method is essentially different from fingerprinting in that our localization method is a \textit{geometric-mapping} rather than fingerprinting\footnote{These terminologies are consistent with the survey in~\cite{yang2013rssi}. Hence, for more details, the readership may be encouraged to refer to~\cite{yang2013rssi}.}. 
More specifically, our proposed localization method directly calculates the location of RF transmitters from  physical measurements (i.e., RSSI and RGB images) with no training datasets.
Hence, we focus on geometric-mapping methods that do not require any training datasets to provide a clear comparison.

\subsection{RSSI-based Localization}
As summarized in Table~\ref{table:related_works_RSSI}, the use of RSSI is a widely applied approach with a long history~\cite{zafari2019survey}.
This RSSI-based localization dates back to 2000, where Microsoft research presented a RADAR~\cite{bahl2000radar} that localizes the RF transmitter via triangulation from the RSSI of three RF anchors.
This could be done because combined with a path-loss model, RSSI gives information on how far the transmitter and receiver are separated.
However, this traditional RSSI-based localization possesses the following a drawback that, to derive the three-dimensional coordinate of the RF transmitter from the RSSI, the transmitter must be covered by more than three RF anchors with feasible received powers, and the RSSI should be accessed from all RF anchors.
This requirement is not necessarily satisfied with the daily use of commercial wireless systems.
Meanwhile, by leveraging the CV technique, we overcome these limitations and realize localization with RSSI from a single AP with a single antenna, hence expanding the availability of this RSSI localization.

It should be noted that several studies ~\cite{li2019cost, kokkinis2019rss, pajovic2019fingerprinting,passafiume2017enhanced} proposed RSSI localization methods with a single RF anchor with decimeter-level accuracy.
However, the common requirement is the implementation of multiple antennas or antenna elements to estimate the angle of arrival (AoA).
Moreover, \cite{li2019cost, kokkinis2019rss, pajovic2019fingerprinting} require fingerprinting databases, and \cite{passafiume2017enhanced} requires multiple bands.
Unlike these studies, we propose a localization method using only single-antenna and single-anchor RSSI with no fingerprinting databases (and a training dataset to construct the databases) and using a single band.

\subsection{CSI-based Localization}
RSSI from a single RF anchor is deficient in localizing an RF transmitter because RSSI provides information only to distances from the target transmitter with coarse granularity, and the AoA cannot be estimated.
Meanwhile, more fine-grained RF information (i.e., CSI) involves more informative features, particularly on multipath characteristics in signal propagation, thus enabling the estimation of the AoA.
This could be done along with the implementation of orthogonal frequency division multiple access (OFDM) in a commercial WiFi AP and availability of subcarrier-wise CSI.
Moreover, recent trends for using multiple antenna elements in WiFi necessitate the capture of antenna-element-wise CSI, and this is informative for estimating the AoA.
Hereinafter, we review CSI-based localization focusing on methods that only require a single RF anchor's CSI.

\subsubsection{Single RF anchor Localization with Multiple Antenna Elements}
Recent WiFi APs compliant with the IEEE 802.11n/ac/ax standards implement multiple antenna elements to perform multiple-input-multiple-output (MIMO) transmission.
The antenna-element-wise CSI obtained therein allows the capture of not only the transmitter/receiver distance but also the AoA, thus enabling transmitter localization.
The work in~\cite{vasisht2016decimeter} derived an antenna element-wise time-of-flight from CSI via sophisticated calibration and estimated the localization of the transmitter.
Similarly, the work in~\cite{wielandt2017indoor} estimated the AoA with CSI from multiple antenna elements, wherein a multipath effect is harnessed by combining a floor plan to accurately localize a transmitter.
The work in~\cite{sen2013avoiding} leveraged the MUSIC algorithm to estimate the AoA from antenna element-wise CSI.
In~\cite{wen2014fine}, a more accurate AoA estimation method was proposed by solving the coherence of CSI among antenna elements.
However, this limits the usage model to a scenario where an AP implements a MIMO transmission mechanism.
Unlike these studies, our proposed localization method can be applied to single-antenna APs, irrespective of the number of implemented antennas.

\subsubsection{Single-Antenna Localization}
With CSI, even from a single antenna element, the RF transmitter can be localized.
This can be achieved by estimating channel decorrelations in a frequency domain that informs multipath characteristics.
From these multipath characteristics, the target can be localized.
The work in~\cite{grosswindhager2018salma} derived the location of the target transmitter from a channel impulse response estimated by ranging in a wide-band measurement and a floor plan.
In~\cite{suraweera2018passive}, the time difference of arrival is estimated from the subcarrier-wise CSI in OFDM symbols, thus localizing and tracking the target transmitter via particle filters.
However, this method of leveraging the channel decorrelations in the frequency domain requires underlying wireless systems to use a wide bandwidth to capture the decorrelations.
Unlike these methods, our proposed method is agnostic to the bandwidth of the underlying wireless systems.

Another CSI-based localization method with a single antenna element is based on the time-domain decorrelation of the transmitter signals (i.e., Doppler shift).
In~\cite{komamiya2019single, komamiya2020radiation}, an AoA was estimated via the receiver's displacement and phase shift of the received signals, and thus, the target transmitter was localized.
However, this was applied to vehicle-to-vehicle communication and cannot be applied to a static environment in which the transmitter and receiver are not moving.
This means to find a static transmitter, a receiver should be displaced (i.e., an RF anchor), which may be cumbersome for a user, particularly in wireless systems operated in a static infrastructure mode.
Unlike these methods, our proposed method can be applied to static scenarios.

\section{System Model}
\begin{figure}[t!]
	\centering
	\includegraphics[width=0.44\textwidth]{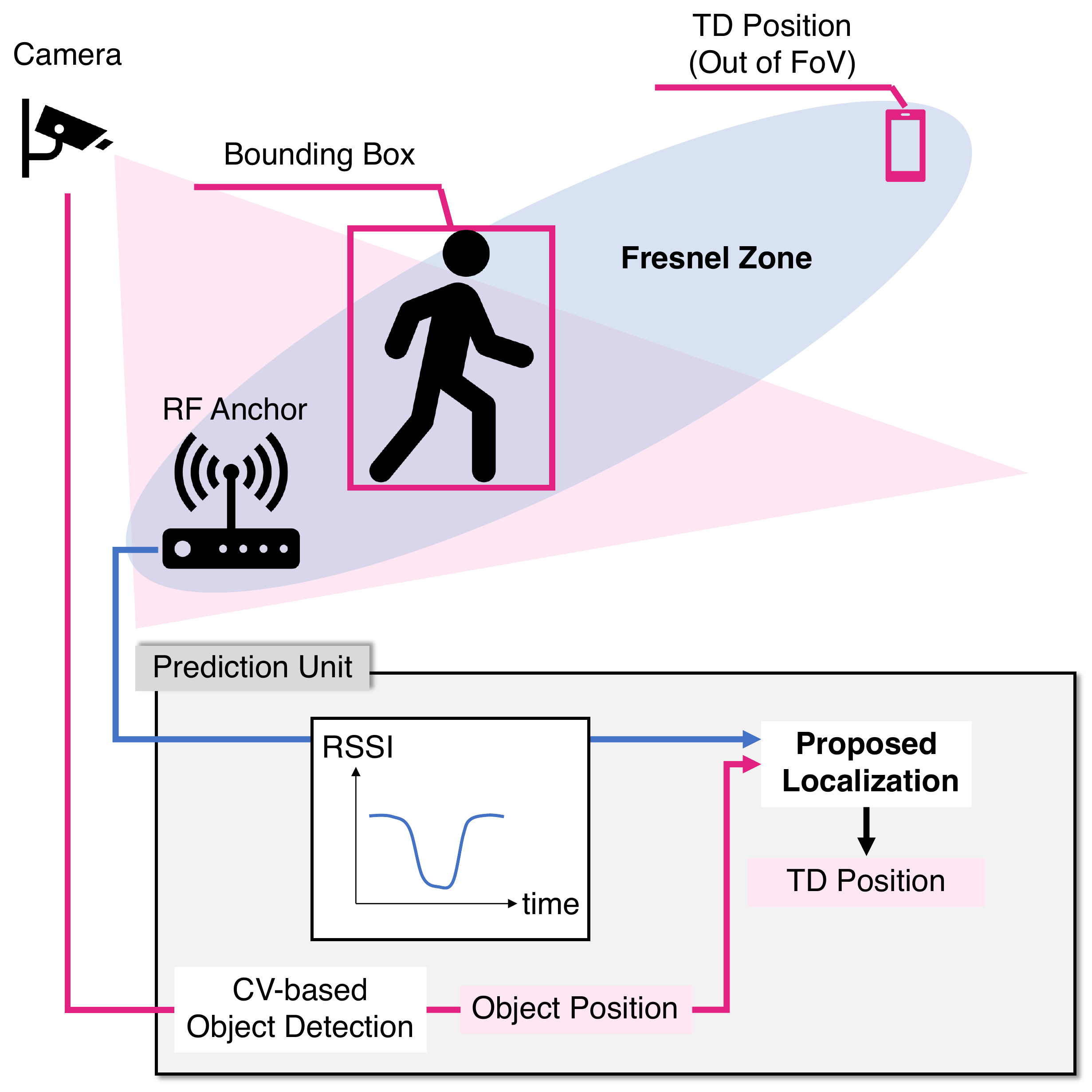}
	\caption{System model.
		The prediction unit estimates the position of the target device based on the camera image and the received power measured at the RF anchor.
		The target device and RF anchor were located on the focus of the Fresnel ellipse.
		Note that the system does not require the camera to detect the target device.
		The procedure for the proposed localization is shown in Figs.~\ref{fig:SAL}, and \ref{fig:fitting}.
	}
	\label{fig:system_model}
\end{figure}

\vspace{.3em}\noindent \textbf{System components.}\quad
As shown in Fig.~\ref{fig:system_model}, the considered system comprises a single RF anchor, an RGB-D camera, a prediction unit, and target transmitter devices (TDs), with the aim of estimating the relative positions of the TDs from the RF anchor.
The RGB-D camera, typified by ZED~\cite{webpage_zed}, captures not only color images but also depth images obtained from a stereo camera.
The RGB-D camera is required to detect nothing but obstacle (discussed below) positions, indicating that the RGB-D camera does not need to capture the positions of the TDs.
This indicates that our system applies to scenarios where TDs exist outside the field of view (FoV) of the RGB-D camera, as shown in Fig.~\ref{fig:system_model}, or they are too small to be detected from the camera images.
The RF anchor measures the received powers of the frames transmitted by the TDs, which is exemplified by beacon frames or uplink data frames.
The RF anchor is not required to capture CSI than the received power.
Therefore, we can leverage not only APs but also terminals that can detect and measure the RSSI of the frames of the TD (e.g., packet sniffer) as the RF anchor.

\vspace{.3em}\noindent \textbf{System requirements.}\quad
In addition to the above components, we leverage a blockage obstacle that blocks the FFZ between the TDs and the RF anchor periodically.
As discussed above, we assume that the RGB-camera can detect obstacles, whereas the camera does not capture TDs.
This applies to most cases because the obstacles (e.g., humans, cars, and robots) are generally much larger than the TDs (e.g., sensors, tags, and smartphones).
Additionally, in order for the received power values to reflect the FFZ blockage caused by the obstacle, the high-frequency RF signals (e.g. 2.4\,GHz, 5\,GHz and mmWave bands) must be transmitted from TDs.
Because, generally, the higher the frequency used by TDs, the greater the attenuation of the received power occurs when the FFZ is blocked,  (i.e., when the line-of-sight path is blocked). 
Particularly, the frequency should be higher than at least 2.4\,GHz~\cite{turner2013human}, which is widely used in wireless LAN and personal area network (PAN) systems (e.g., Wi-Fi, Bluetooth, and ZigBee).

\vspace{.3em}\noindent \textbf{Localization flow.}\quad
The prediction unit is connected to the RF anchor and RGB-D camera, and performs localization as follows:
First, the RF anchor and RGB-D camera send the time series of the received powers and RGB-D images, respectively, to the prediction unit.
Subsequently, based on the received RGB-D images, the prediction unit estimates obstacle regions using a CV-based object detection algorithm, such as YOLO~\cite{redmon2016you}.
Generally object detection algorithms output the bounding box of the obstacle, which is a rectangle enclosing an object in the image as shown in Fig.~\ref{fig:system_model}.
From the bounding box, the obstacle region is obtained, which is detailed in Section~\ref{ssec:FZB_estimation}.
Finally, by referring to both the historical obstacle region and received powers, the prediction unit estimates the ``blockage position,'' referred to as the nearest point in the obstacle region to the FFZ where the blockage event starts or ends.
Based on this, TD is localized as discussed in the subsequent section.

\section{Proposed RSSI Localization Method}
Our proposed SARR-LOC leverages dynamic blockage events, thus estimating the FFZ between each TD and the RF anchor.
As show in the existing works~\cite{rampa2015physical,obayashi1998body}, when obstacle overlaps with the FFZ, the received power is sharply decreased.
We capitalized on this fact by hypothesizing that by gathering information on the region of the obstacle at which the received power decreases, we can obtain several points of the FFZ boundary; thus, we can estimate the entire FFZ.
Note that from the FFZ, the TD position (i.e., the distance of the TD and the RF anchor $d$ and the angle of the TD to the RF anchor $\theta$) can be determined uniquely, and vice versa; hence, the estimations of the FFZ and TD positions are interchangeable.

More specifically, based on the Fresnel zone model~\cite{molisch2012wireless,monk2010fresnel}, the FFZ boundary generally forms an ellipse parameterized by $d$ and $\theta$; hence, by estimating the most likely $d$ and $\theta$ given several points on the FFZ boundary, we can localize the TD position.
Mathematically, given $d$, $\theta$, and the wavelength $\lambda$,
the FFZ boundary is represented~\cite{monk2010fresnel} as the following ellipse equation:
\begin{align}
	\label{equ:ellipse}
	F & (x,y,d,\theta) \notag                                                                                                                       \\
	  & = \frac{(4x \cos \theta + 4y \sin \theta  -2d)^2}{(2d + \lambda)^2}  +\frac{(4x \sin \theta - 4y \cos \theta)^2}{\lambda(4d+\lambda)}\notag \\
	  & =1,
\end{align}
where $(x, y)$ denotes the arbitrary points on the ellipse.
Based on this function, we estimate the most likely $d$ and $\theta$ based on the set of estimated FFZ boundary points.

\begin{figure}[t!]
	\centering
	\includegraphics[width=0.47\textwidth]{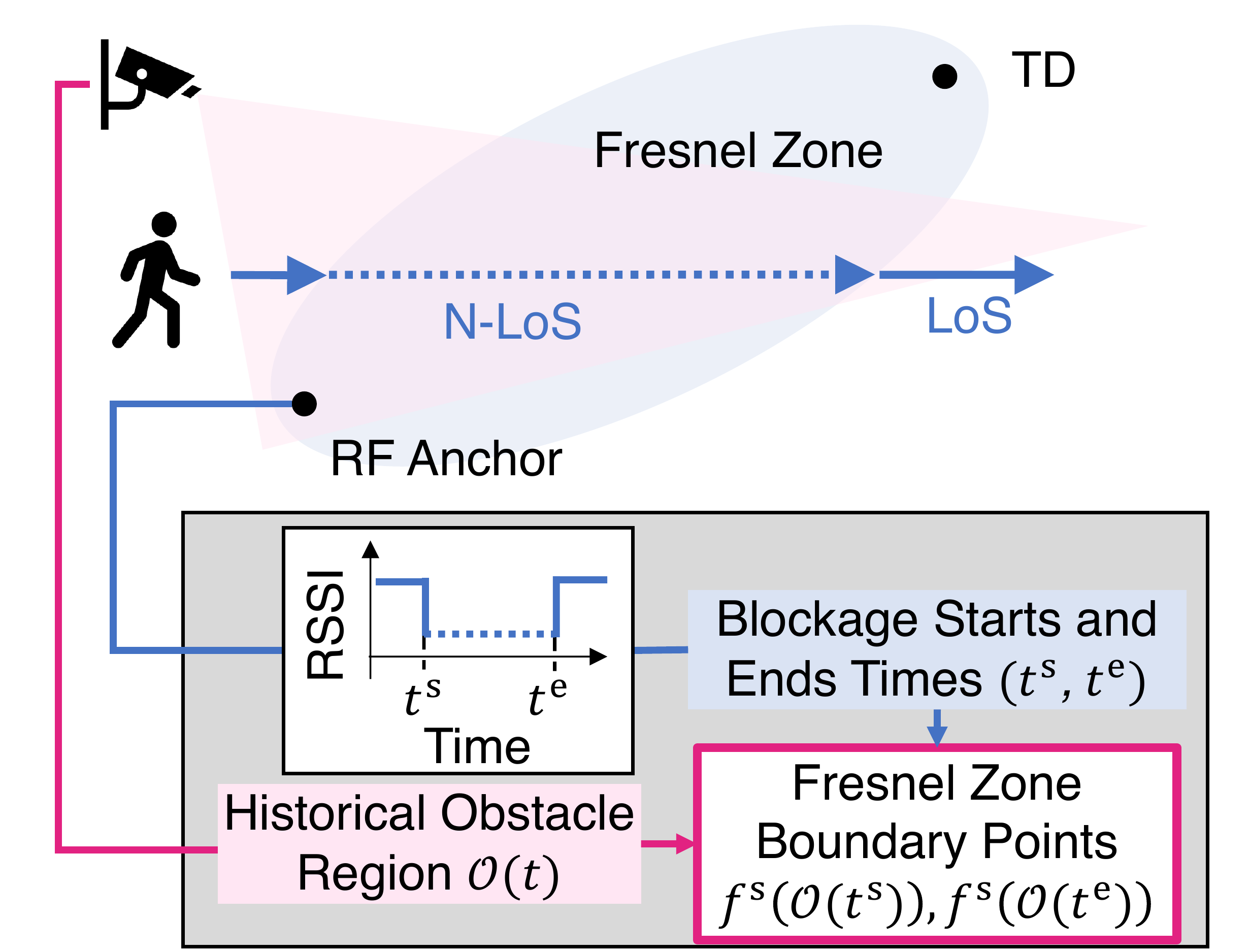}
	\caption{First Fresnel zone boundary points estimation.
		Using the blockage start and end times $t^\mathrm{s}, t^\mathrm{e}$, and the historical obstacle region $\mathcal{O}(t)$,
		the first Fresnel zone boundary points is estimated as: $f^\mathrm{s}(\mathcal{O}(t^\mathrm{s}))$ and $f^\mathrm{s}(\mathcal{O}(t^\mathrm{e}))$, where $f^\mathrm{s}(\cdot)$ and $f^\mathrm{e}(\cdot)$ indicate the first Fresnel zone boundary estimation function.
	}
	\label{fig:SAL}
\end{figure}

\begin{figure}[t!]
	\centering
	\subfloat[First Fresnel zone boundary points.]{
		\includegraphics[width=0.22\textwidth,page =1]{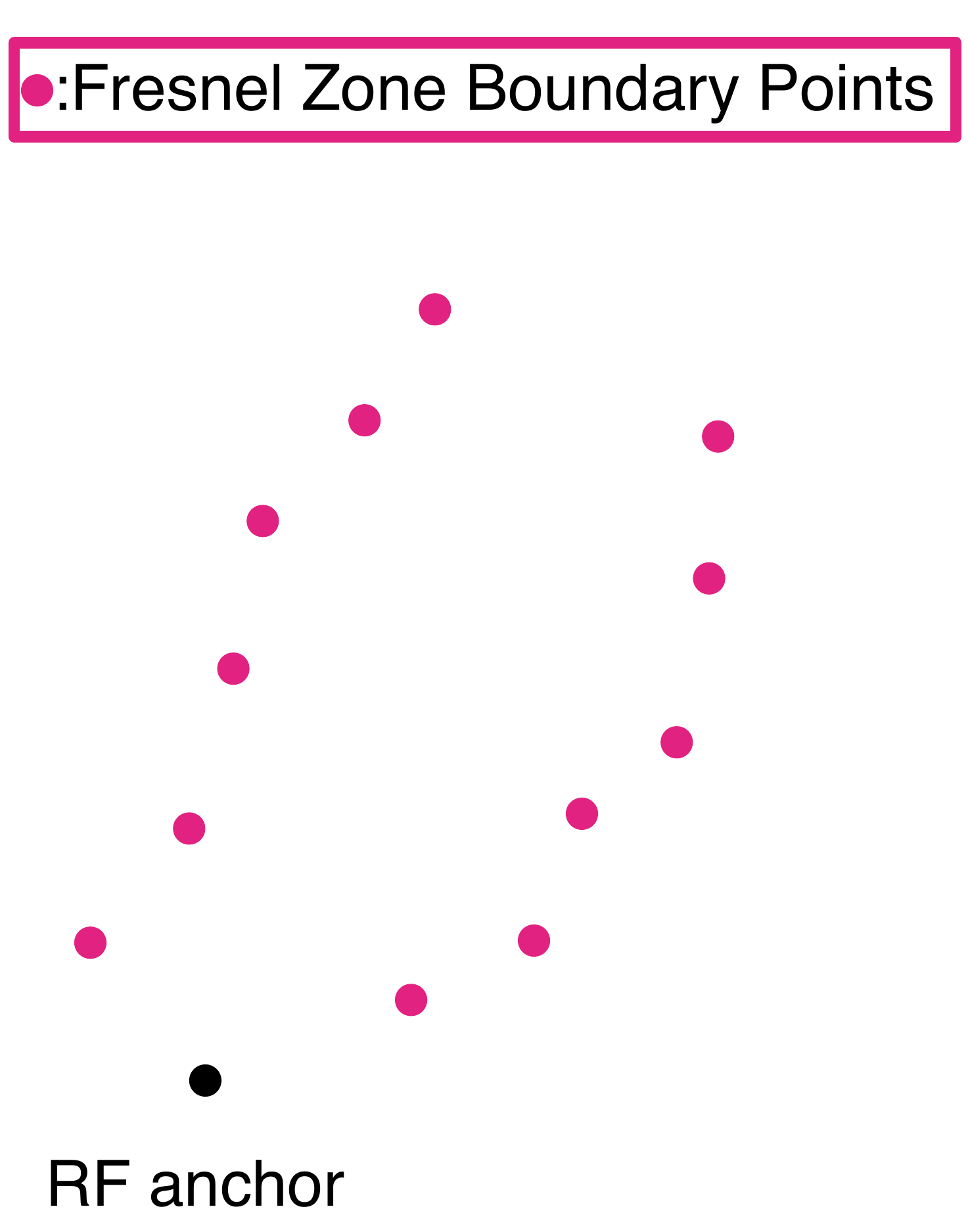}}
	\subfloat[First Fresnel ellipse fitting.]{
		\includegraphics[width=0.22\textwidth,page =2]{Fitting.pdf}}
	\caption{First Fresnel ellipse fitting.
		The first Fresnel equation \eqref{equ:ellipse} is fitted to the first Fresnel zone boundary points, as shown in Fig.~\ref{fig:SAL}.
		Then we estimate the TD position.
	}
	\label{fig:fitting}
\end{figure}

Hence, the problem boils down to estimating several points on the FFZ boundary and deriving the most likely $d$ and $\theta$, each of which is depicted in Figs.~\ref{fig:SAL} and \ref{fig:fitting}, respectively.
As shown in Fig.~\ref{fig:SAL}, the points on the FFZ boundary are obtained using the time series of the received power measured at the RF anchor and the historical region of the obstacle, harnessing the fact~\cite{rampa2015physical} that the received power sharply decreases when the obstacle reaches the FFZ boundary.
Next, given the several estimated points on the FFZ boundary, as shown in Fig.~\ref{fig:fitting}, the parameters $d$ and $\theta$ were estimated by fitting \eqref{equ:ellipse}.
The detailed procedures are elaborated in the following sections.

\subsection{First Fresenel Zone Boundary Point Estimation from Received Power and Obstacle Position}
\label{ssec:fresnel_zone_estimation}
To obtain the points on the FFZ boundary, we harness the relationship between the position of the obstacle and the received power.
In more detail, the points are estimated from the obstacle regions at the blockage event starts and ends,
which are estimated as the time when the degrade of the received power starts or ends, respectively.
This is derived by the FFZ blockage model~\cite{rampa2015physical} that when the obstacle overlaps to the FFZ, the received power is decreased, which is referred to as the blockage event.
The blockage event is defined as temporal attenuation of the received power that the received power decreases, remains low level for a longer time than that the obstacle to move by its size\footnote{Temporal attenuation is also caused by factors other than FFZ blockage, such as noises and fading.
	Comparing to the temporal attenuations by the noises and fadings, the received power in the temporal attenuation by FFZ blockage remains low level for a longer time length.
	Based on this, the blockage event is distinguished from the temporal attenuation by the noises or fadings in terms of the time length of remaining at a low level.}, and then increases.
The blockage event starts time $t^\mathrm{s}$ and ends time $t^\mathrm{e}$ are defined as the time when the received power starts to decrease and ends to increase, respectively.
The time $t^\mathrm{s}$ and $t^\mathrm{e}$ are estimated using the blockage detection method from the time series of the measured received power at the RF anchor, which is detailed below.
The points on the FFZ boundary are obtained as the nearest point in the object region at time $t^\mathrm{s}$ and $t^\mathrm{e}$ to the FFZ, which is described in \ref{ssec:FZB_estimation}.

To estimate the blockage start time $t^\mathrm{s}$ and the blockage end time $t^\mathrm{e}$ from the time series of the received power measured at the RF anchor, this study examines a template matching method~\cite{brunelli2009template}.
Algorithm~\ref{alg:blockage} shows the procedure for the blockage start and end times estimation.
In the template matching method, the time required to achieve a high correlation of the template and measured received power is considered to be the start time $t^\mathrm{s}$ for the blockage event.
The end time of the blockage event $t^\mathrm{e}$ was estimated by adding the template time length to the blockage start time $t^\mathrm{s}$.

\begin{figure}[t!]
	\centering
	\includegraphics[width=0.4\textwidth]{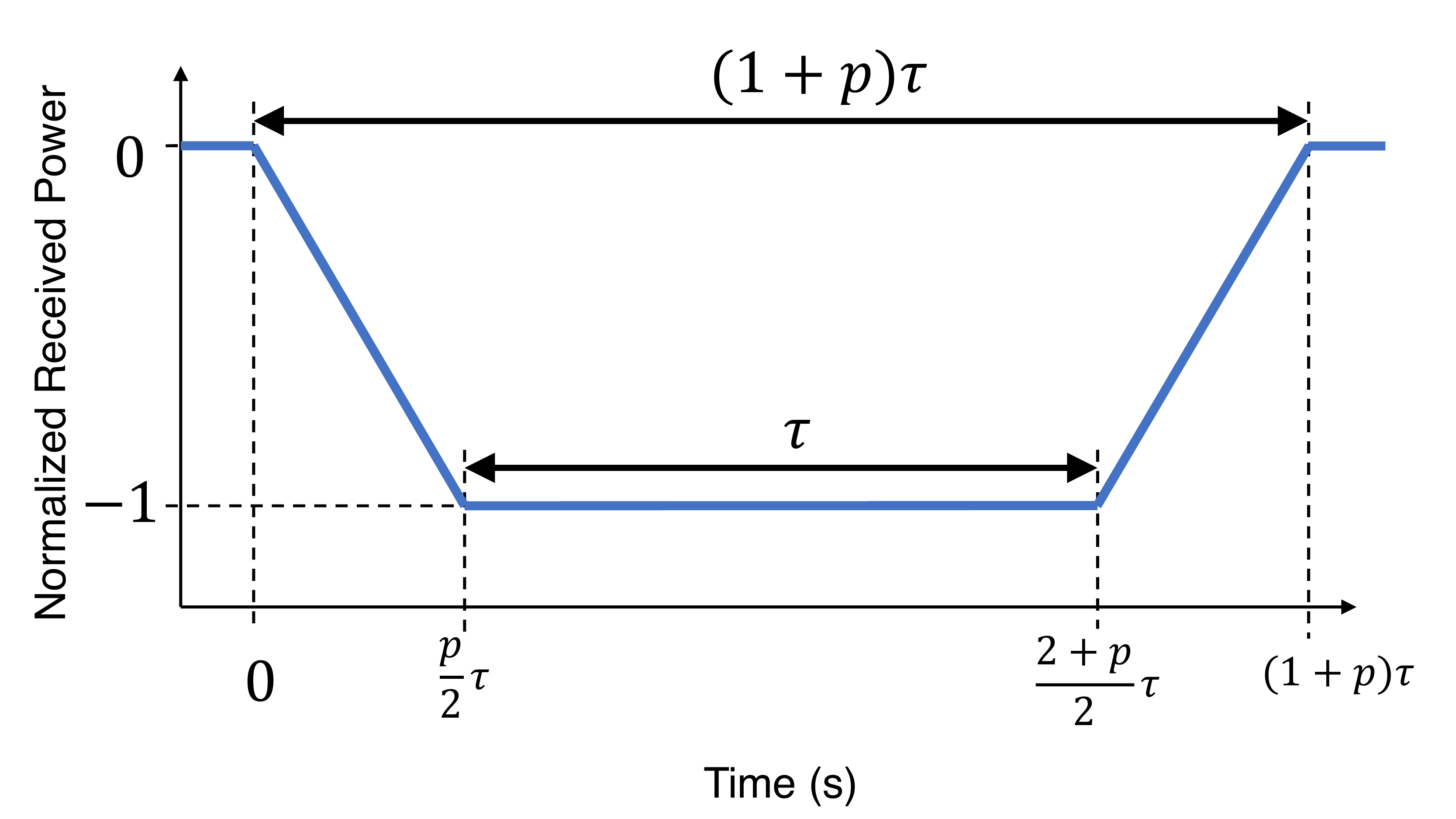}
	\caption{Blockage detection template, which indicates the time-varying received power model of the dynamic blockage event.}
	\label{fig:temperate}
\end{figure}

As the blockage template, we examined a simple piecewise linear function~\cite{jacob2010adynamic}, as illustrated in Fig.~\ref{fig:temperate}.
Such simple modeling is based on the existing human body blockage evaluations~\cite{obayashi1998body}, moreover, we experimentally validated that the simple blockage template accurately estimates the Fresnel zone (see Table~\ref{tab:miss_detection}, Fig~\ref{fig:blockage_detection_result}, and Fig.~\ref{fig:fresnel_point_estimation}~(a)).
This template function $r^\mathrm{tmp}(t,\bm{w})$ is denoted as:
\begin{align}
	\label{equ:filter}
	r^\mathrm{tmp}(t,\bm{w}) & =
	\begin{cases}
		- \frac{2}{p\tau}t              & 0<t< \frac{p}{2}\tau;                                                         \\
		-1                              & \frac{p\tau}{2}<t<\frac{2+p}{2}\tau;                                          \\
		\frac{2}{p\tau}t-\frac{2+2p}{p} & \frac{2+p}{2}\tau<t<(1+p)\tau;                                                \\
		0                               & \text{otherwise},                                                            
	\end{cases}                                   \\
	\bm{w}                   & \coloneqq [p, \tau]^{\mathrm{T}},
\end{align}
where $t=0$ indicates the blockage event starts time and $\bm{w}$ indicates a parameter vector $[p, \tau]^{\mathrm{T}}$.
Therein, the normalized received power of $-1$ and $0$ indicate mean blockage power and power without any blockage  respectively.
The template time length $(1+p)\tau$ indicates the duration between the last zero crossing before and the first zero crossing after the blockage event.
The decay and rising time $p/2\tau$ specify the time span between the zero crossings of the signal level and a  mean blockage power.

To detect various blockage events flexibly, multiple templates with different parameters were used, and thus, we select the most likely blockage detection from the detection of multiple templates for each blockage event.
This is because the blockage time length differs from the width of the FFZ, which depends on the distance between the TD and the RF anchor and the velocity of the obstacle.
This procedure is elaborated in Section~\ref{subsec:template_param_adaptation}.

\begin{algorithm}[t]
	\caption{Procedure of the blockage stars and ends time estimation}
	\label{alg:blockage}
	\begin{algorithmic}[1]
		\Mydef SARR-LOC
		\Input $\bm{p}(t), r(t)$
		\State $\bm{t}^\mathrm{s}, \bm{t}^\mathrm{e} = \text{BlockageDetection}(r(t))$
		\State  $\bm{P} = \{f^\mathrm{s}(\mathcal{O}(t)) \mid t \in \bm{t}^\mathrm{s} \} \cup \{f^\mathrm{e}(\mathcal{O}(t)) \mid t \in \bm{t}^\mathrm{e} \}$.
		\State $(d^\star, \theta^\star) = \argmin_{(d, \theta)} \sum_{(x, y) \in \bm{P}} [F(x,y,d,\theta)-1]^2$
		\Myreturn{$d^\star,\theta^\star$}
		\\
		\Mydef BlockageDetection
		\Input $r(t),\bm{W}=\{\bm{w}_1,\bm{w}_2,\dots,\bm{w}_K\}$
		\State $\bm{t}^\mathrm{s}=\emptyset, \bm{t}^\mathrm{e}=\emptyset$
		\For {Each parameter $\bm{w}_k$}
		\State $\bm{c} = \mathrm{NormalizedCorrelation}(r(t),r^\mathrm{tmp}(t,\bm{w}_k))$
		\State $\mathcal{T}^\mathrm{peak}\times\mathcal{C}^\mathrm{peak} = \mathrm{PeakDetection}(\bm{c})$
		\State $\mathcal{T}^\mathrm{s}_k\times\mathcal{C}^\mathrm{s}_k = \{(t, c) \mid (t, c) \in \mathcal{T}^\mathrm{peak}\times\mathcal{C}^\mathrm{peak}, c > c^\mathrm{th}\}$
		\State $\mathcal{T}^\mathrm{s}_k\times\mathcal{C}^\mathrm{s}_k$ are vectorized to $|\mathcal{T}^\mathrm{s}_k|\times 2$ matrix $(\bm{T}^\mathrm{s}_k,\bm{C}^\mathrm{s}_k)$
		\For {$1\leq i\leq |\mathcal{T}^\mathrm{s}_k|$}
		\State $T^\mathrm{e}_{k,i} = T^\mathrm{s}_i+T^\mathrm{tmp}$
		\EndFor
		\State $\mathcal{T}^\mathrm{sec}_{k} = \{t \mid t^\mathrm{s}_i < t < t^\mathrm{e}_i,1 \leq i \leq |\mathcal{T}^\mathrm{s}_k| \}$
		\EndFor
		\State $t_0 = 0$
		\While {$t_0<t^\mathrm{obs}$}
		\State $\mathcal{K}\times\mathcal{M}=\{(k, m) \mid T^\mathrm{s}_{k,m}<t_0 < T^\mathrm{e}_{k,m}\}$
		\If {$\mathcal{K} = \emptyset$}
		\State $t_0= t_0 + \tau^\mathrm{s},$ and continue;
		\EndIf
		\State $(k^\star,m^\star) = \mathop{\rm arg~max}\limits_{(k,m) \in \mathcal{K}\times\mathcal{M}} C^\mathrm{s}_{k,m}.$
		\State $T^\mathrm{s}_{k^\star,m^\star},T^\mathrm{e}_{k^\star,m^\star}$ are appended to  $\bm{t}^\mathrm{s},\bm{t}^\mathrm{e}$, respectively.
		\State $t_0= T^\mathrm{e}_{k^\star,m^\star}$
		\EndWhile
		\Myreturn $\bm{t}^\mathrm{s}, \bm{t}^\mathrm{e}$
	\end{algorithmic}
\end{algorithm}

\subsubsection{Implementation of Blockage Detection from the Received Power}
\label{ssec:blockage_detection}
This section describes the detailed procedure for blockage detection using a blockage template parameter $\bm{w}$.
First, a zero-mean normalized correlation function $c(t)$ of the time-varying received power $r(t)$ and a blockage template $r^\mathrm{tmp}(t,\bm{w})$ are calculated.
Subsequently, the time required to achieve a high correlation function is considered to be the blockage starting time, because the time $0\,\mathrm{s}$ of the blockage template indicates the blockage event start time.

In more detail, the correlation series $c_i$ is calculated as:
\begin{align}
	\label{equ:correlation}
	c_i                    & =\notag \frac{\sum_{j=0}^{n^\mathrm{tmp}} 
	\bar{r}_{i,j} \bar{r}^\mathrm{tmp}_j}
	{\sqrt{\sum_{j=0}^{n^\mathrm{tmp}}  \{\bar{r}_{i,j}\}^2 
	\sum_{j=0}^{n^\mathrm{tmp}}  \{\bar{r}^\mathrm{tmp}_j\}^2 }},                                                                                                            \\
	\bar{r}_{i,j}          & \coloneqq r(i\tau^\mathrm{s}+j\tau^\mathrm{s})- \frac{1}{n^\mathrm{tmp}+1} \sum_{k=0}^{n^\mathrm{tmp}}  r(i\tau^\mathrm{s} + k\tau^\mathrm{s}), \\
	\bar{r}^\mathrm{tmp}_j & \coloneqq r^\mathrm{tmp}(j\tau^\mathrm{s},\bm{w}) - \frac{1}{n^\mathrm{tmp}+1}
	\sum_{k=0}^{n^\mathrm{tmp}} r^\mathrm{tmp}(k\tau^\mathrm{s},\bm{w}),
\end{align}
where $r(t)$ and $r^\mathrm{tmp}(t,\bm{w})$ are sampled with sampling interval $\tau^\mathrm{s}$, 
the template time length is $t^\mathrm{tmp}$, 
and $n^\mathrm{tmp} \coloneqq \lfloor t^\mathrm{tmp}/\tau^\mathrm{s}\rfloor$.
In this study, $t^\mathrm{tmp}$ becomes $(1+p)\tau$, as discussed in the previous section.
For shorthand notation, $\bm{c}$ denotes the concatenation of $\{c_i\}_{i=1}^{n^\mathrm{obs}}$, 
where $t^\mathrm{obs}$ indicates the time length of the received power observation, and $n^\mathrm{obs} \coloneqq \lfloor t^\mathrm{obs}/\tau^\mathrm{s}\rfloor$.
Next, the blockage start times $\mathcal{T}^\mathrm{s}$ are obtained as the time points for achieving high correlation values $\bm{c}$ using peak detection and threshold cut-off.
The peak points of $\bm{c}$ are selected as $\mathcal{C}^\mathrm{peak}$, which achieves the largest correlation in the range of $n^\mathrm{tmp}$, as follows:
\begin{align}
	\label{equ:peak_selection}
	\mathcal{T}^\mathrm{peak} \times \mathcal{C}^\mathrm{peak} = \left\{(i\tau^\mathrm{s}, c_i)  \mid c_i = \max \mathcal{C}^\mathrm{range}_i  \right\},
\end{align}
where 
\begin{align}
	\mathcal{C}^\mathrm{range}_i = \left\{c_j \,\middle|\, i-\frac{n^\mathrm{tmp}}{2}<j<i+\frac{n^\mathrm{tmp}}{2} \right\}.
\end{align}
To cut the noisy peak points, a threshold-based selection is performed and the blockage event start times $\bm{t}^\mathrm{s}$ and its corresponding correlation values as:  
\begin{align}
	\label{equ:candidate_selection}
	\mathcal{T}^\mathrm{s}\times \mathcal{C}^\mathrm{s}  = 
	\{(t,c) \mid (t,c) \in \mathcal{T}^\mathrm{peak} \times \mathcal{C}^\mathrm{peak} \cap  c > c^\mathrm{th} \},
\end{align}
where $c^\mathrm{th}$ indicates a threshold of the correlation.
In the experimental evaluation of this study, $c^\mathrm{th}$ is fixed to 0.6.
We vectorized $\mathcal{T}^\mathrm{s}\times\mathcal{C}^\mathrm{s}$ to $|\mathcal{T}^\mathrm{s}|\times 2$ vector $(\bm{t}^\mathrm{s},\bm{c}^\mathrm{s})$.
Using each blockage start time point $\bm{t}^\mathrm{s}$, the blockage end time points $\bm{t}^\mathrm{e}$, and the blockage time sections $\mathcal{T}^\mathrm{sec}$ are denoted as follows:
\begin{align}
	t^\mathrm{e}_i           & = t^\mathrm{s}_i + t^\mathrm{tmp},                                                                \\
	\mathcal{T}^\mathrm{sec} & = \{t \mid t^\mathrm{s}_i \leq t \leq t^\mathrm{e}_i , 1 \leq i \leq |\mathcal{T}^\mathrm{s}| \}.
\end{align}
Note that the higher the blockage starts correlation, the stronger the estimates that the blockage starts at the corresponding time.
We denote the blockage start times, blockage end times, blockage correlation, and blockage section using the parameter $\bm{w}$ as 
$\bm{t}^\mathrm{s}(\bm{w}),\bm{t}^\mathrm{e}(\bm{w}), \bm{c}^\mathrm{s}(\bm{w}),$ and $\mathcal{T}^\mathrm{sec}(\bm{w})$.

\subsubsection{Multi-Template Based Blockage Detection}
\label{subsec:template_param_adaptation}
To detect various blockage events, multiple templates with different parameters are used to detect blockages and hence,
we select the most likely blockage detection from the detection of multiple templates for each blockage event.
First, the blockage events are detected using multiple templates with different parameter values.
Next, for each blockage event, we determined the most likely detection that achieved a higher correlation than that of the templates with other parameters.
Assuming the parameter searching range as $\{\bm{w}_1,\dots,\bm{w}_K \}$.
For shorthand notation, $\bm{T}^\mathrm{s},\bm{T}^\mathrm{e},$ and $\bm{C}^\mathrm{s}$  are denoted as 
$\{\bm{t}^\mathrm{s}(\bm{w}_k)\}_{k=1}^K,\{\bm{t}^\mathrm{e}(\bm{w}_k)\}_{k=1}^K,$ and $\{\bm{c}^\mathrm{s}(\bm{w}_k)\}_{k=1}^K,$, respectively.

For the detected blockage events, we determine a set of the most likely detections (i.e., $t^\mathrm{s}$ and $t^\mathrm{e}$) that achieves higher correlation than that of the templates with other parameters.
In case of the parameters $\{\bm{w}_{k_1}, \bm{w}_{k_2},\dots,\bm{w}_{k_l}\}$ detect the same blockage event that occur at a time $t_0$.
Formally, the following condition is satisfied  
\begin{align}
	T^\mathrm{s}_{k_j, m_j} < t_0  < T^\mathrm{e}_{k_j, m_j},
\end{align}
where $k_j$ indicates the parameter index $k_j \in \{1,2,\dots,K\}$ and $m_j$ indicates the blockage index of $\bm{T}^\mathrm{s}_{k_j}$, which includes the time $t_0$.
Here we denote:
\begin{align}
	\mathcal{K} \times \mathcal{M}=\{(k,m) \mid T^\mathrm{s}_{k_j, m_j} < t_0  < T^\mathrm{e}_{k_j, m_j}\}.
\end{align}
We select the most likely blockage start and end times for each blockage section as:
\begin{align}
	t^\mathrm{s} = T^\mathrm{s}_{k^\star,m^\star}, \\
	t^\mathrm{e} = T^\mathrm{e}_{k^\star,m^\star},
\end{align}
where 
\begin{align}
	(k^\star,m^\star) = \mathop{\rm arg~max}\limits_{(k,m) \in \mathcal{K} \times \mathcal{M}} C^\mathrm{s}_{k,m}.
\end{align}
The procedure below is iterated for all blockage events detected with at least one template.
Moreover, given the determined set of $t^\mathrm{s}$ and $t^\mathrm{e}$, the FFZ boundary points are estimated as the nearest point in the object region to the FFZ at the time set, which is denoted in the following section.

\subsubsection{First Fresnel Zone Boundary Points Estimation from Blockage Starts and Ends Times.}
\label{ssec:FZB_estimation}
The FFZ boundary points are obtained from the obstacle region at the timings when blockage starts or ends. 
The obstacle region is calculated from the location of the center pixel of the bounding box,
which is denoted as $\bm{p}^\mathrm{cnt}(t)$. 
To simplify the procedure of FFZ boundary points estimation,
we assume that the obstacle is a thin board that the length of the obstacle in the direction perpendicular to the moving direction of the obstacle is ignorable.
Based on the assumption, the obstacle region is considered to be a line segment in 2D space that is parallel to the moving direction.
Formally, given the unit vector that indicates the moving direction of the obstacle as $\bm{n}(t)$ and the width of the obstacle (i.e., the length of the line segment) as $w$,
the obstacle region at time $t$ is denoted as:
\begin{align}
	\mathcal{O}(t) = \left\{\bm{p}^\mathrm{cnt}(t) + k\bm{v}(t) \relmiddle| |k| < \frac{w}{2}\right\}.
\end{align}
The two edge points of the obstacle region are denoted as $\bm{e}^\mathrm{+}$ and $\bm{e}^\mathrm{-}$, where
\begin{align}
	\bm{e}^\mathrm{+}(t) = \bm{p}^\mathrm{cnt}(t) + \frac{w}{2}\bm{n}(t), \\
	\bm{e}^\mathrm{-}(t) = \bm{p}^\mathrm{cnt}(t) - \frac{w}{2}\bm{n}(t).
\end{align}

The nearest point of the obstacle region to the FFZ depends on whether the blockage starts or ends.
The mathematical explanation of the following procedure is detailed in Appendix. 
When the blockage event starts, this implies that the obstacle enters to the FFZ.
Thus, the nearest point is the furthest point from the center point in direction of the obstacle movement, which results that the nearest point is $\bm{e}^\mathrm{+}$.
On the other hand, when the blockage event ends, which implies that the obstacle leaves from the FFZ.
Thus, the nearest point is the furthest point from the center point in reverse direction of the obstacle movement, which results that the nearest point is $\bm{e}^\mathrm{-}$.
For shorthand notation, the FFZ boundary point estimation function at the blockage starts time  and the blockage ends time are respectively defined as follows:
\begin{align}
	f^\mathrm{s}(\mathcal{O}(t^\mathrm{s})) = \bm{e}^\mathrm{+}(t^\mathrm{s}), \\
	f^\mathrm{e}(\mathcal{O}(t^\mathrm{e})) = \bm{e}^\mathrm{-}(t^\mathrm{e}).
\end{align}

\subsection{Frist Fresenel Ellipse Fitting}
To estimate the parameters of \eqref{equ:ellipse} from the points on the FFZ boundary, as described in Section~\ref{ssec:blockage_detection}, 
we fit \eqref{equ:ellipse} to the FFZ boundary points using the least squares method.
In the least squares method, we define a loss function as 
\begin{align}
	L(x,y,d,\theta)= [F(x,y,d,\theta)-1]^2,
\end{align}
where $(x,y)$ indicates the FFZ boundary point and $F(\cdot)$ is denoted by \eqref{equ:ellipse}.
Subsequently, to estimate the parameters, we solve the optimization problem as follows:
\begin{align}
	\label{equ:optimization}
	(d^{*}, \theta^{*}) = \argmin_{(d, \theta)} \sum_{(x, y) \in \bm{P}} L(x,y,d,\theta),
\end{align}
where 
\begin{align}
	\bm{P} =  \{f^\mathrm{s}(\mathcal{O}(t)) \mid t \in \bm{t}^\mathrm{s} \} \cup \{f^\mathrm{e}(\mathcal{O}(t)) \mid t \in \bm{t}^\mathrm{e} \}.
\end{align}
This study examines a grid search method to solve the optimization problem. 
The detailed method to solve \eqref{equ:optimization} is denoted in Appendix~\ref{ssec:deatil:fitting}.

\section{Experimental Evaluation}
\subsection{Setup}
\begin{figure}[t!]
	\centering
	\subfloat[(a) Snapshot of experimental setup.]{
		\includegraphics[width=0.4\textwidth]{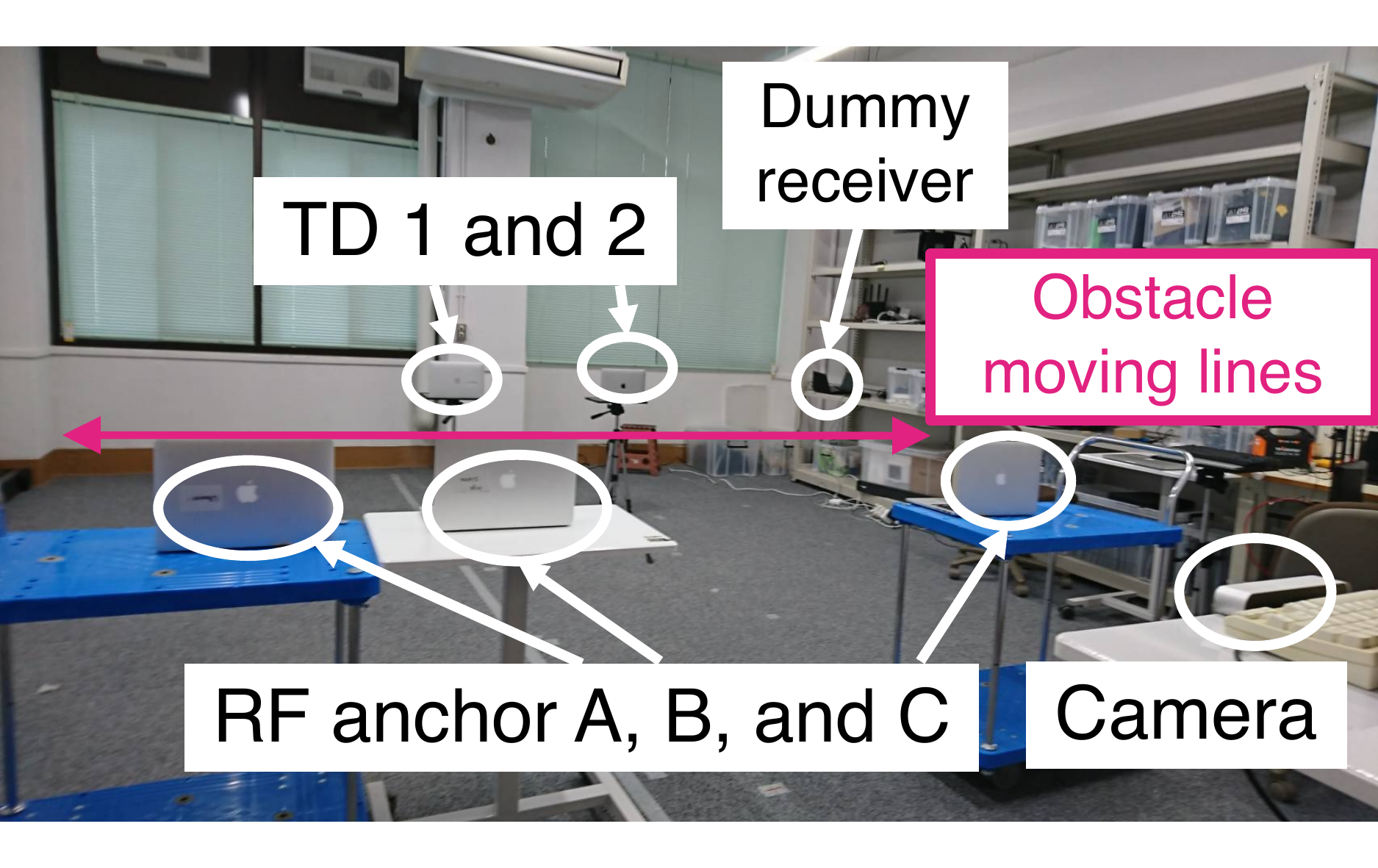}
		\label{fig:experiment_photo}
	}\\
	\subfloat[(b) Equipments deployment.]{
		\includegraphics[width=0.4\textwidth]{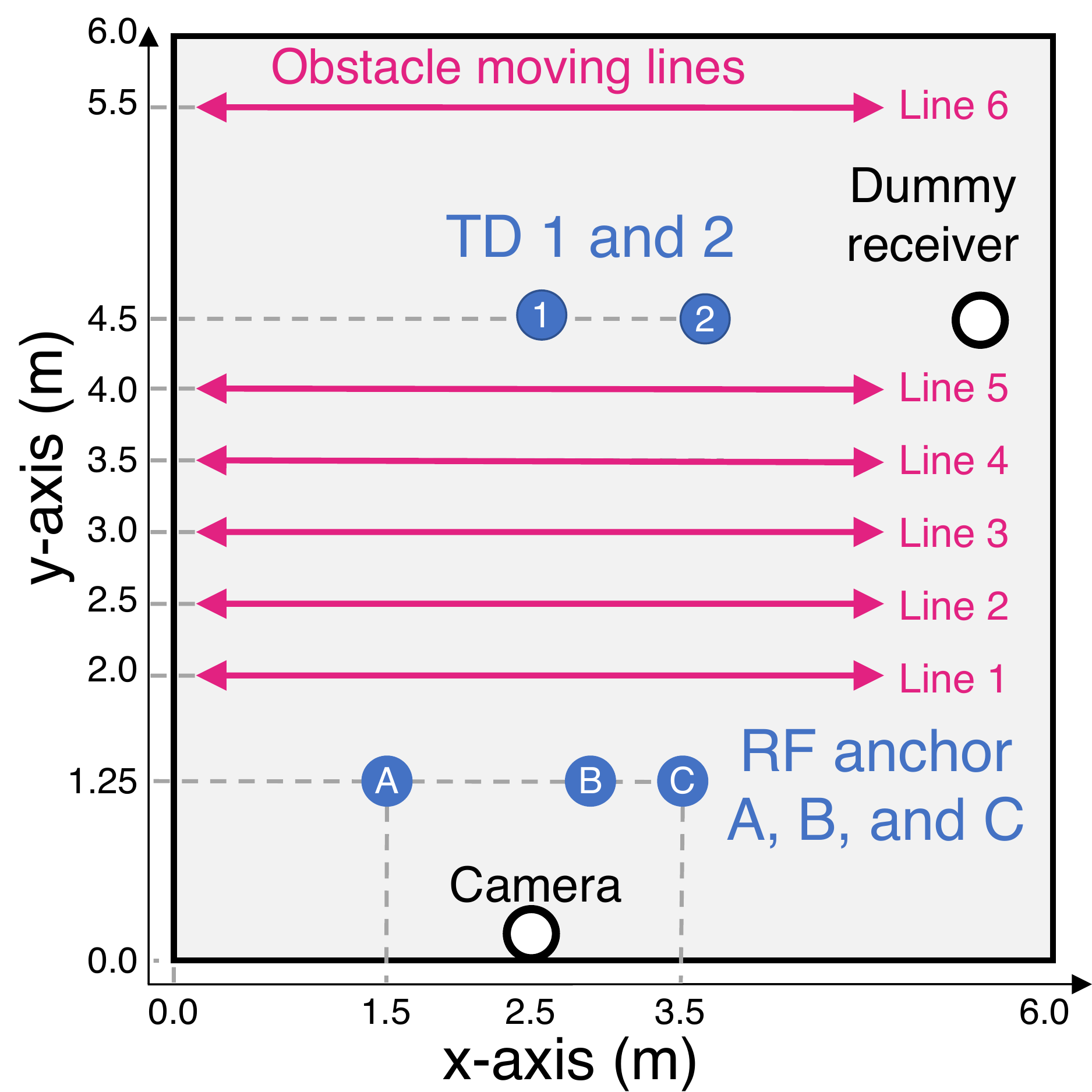}
	}
	\caption{Experimental setup.
		The room has a height of 3.8\,m, width of 6.0\,m, and depth of 6.0\,m.
		The black square indicates the concrete wall. 
		The dummy receiver, TDs, RF anchors, and camera were at height of 0.75\,m.}
	\label{fig:setup}
\end{figure}

The experimental setup is illustrated in Fig~\ref{fig:setup}.
We deployed target TDs, RF anchors, an RGB-D camera (i.e., ZED~\cite{webpage_zed}), and a dummy receiver.
The TDs were connected to the network via a dummy receiver and sent packets to the dummy receiver using iPerf~\cite{iperf}.
The frames from the TDs to the dummy receiver were captured and measured at the RF anchors.
The locations of the dummy receiver, TDs, RF anchors, and camera are shown in Fig.~\ref{fig:setup}~(a), where they were placed at heights of 0.75\,m.
Details of the experimental equipment are provided in Table~\ref{table:equipment}.
The experimental environment was an indoor room with a height of 3.8\,m, width of 6.0\,m, and depth of 6.0\,m, where the material of the wall was concrete.

The obstacle was a steel panel with a height of 1.8\,m and a width of 0.9\,m,
which is moved by a human with an average speed of 0.57\,$\mathrm{m/s}$ and blocked the FFZ between the TDs and the RF anchors.
Note that the width of the obstacle is larger than the width of the FFZ.
As illustrated in Fig.~\ref{fig:setup}, the moving lines of the obstacle are along with either of the six lines, line 1 to 6, lying parallel to the x-axis.
The obstacle took six round trips for each line.
Using the RGB-D camera image and a machine-learning-based object detection algorithm, YOLO~\cite{redmon2016you}, the center positions of the human were detected,
where the position of the human was treated as the center positions of the obstacle.
Moreover, the positions of the RF anchors were localized by object detection beforehand.
The experiment was performed for approximately 13 minutes.
\begin{table}[t]
	\caption{Experimental equipment.}
	\label{table:equipment}
	\centering
	\scalebox{0.90}[0.90] {
		\renewcommand\arraystretch{1.1}
		\begin{tabular}{cc}
			\toprule
			TDs and RF anchors & MacBook Pro            \\
			Dummy receiver     & NETGEAR Nighthawk AX12 \\
			RGB-D camera       & ZED~\cite{webpage_zed} \\
			Protocol           & IEEE 802.11ac          \\
			Wireless Band      & 5.22\,GHz, 44\,ch      \\
			\bottomrule
		\end{tabular}
	}
\end{table}

Owing to the error in the depth estimation by the ZED stereo camera, the CV-based object detection system includes an error in the localization of the obstacles.
To correct this, the object detection results were calibrated using a simple linear regression model.
To develop the model, we created a dataset comprising the actual measured human positions and the center position of the human estimated by the CV system.
The training dataset contains 16 points around the room.
The calibration model is trained to estimate the ground-truth position from the position output by the CV system, thus correcting the CV positioning errors.
Moreover, evaluating the error of depth estimation and adapting more sophisticated calibration methods are beyond our scope, 
because our objective is to show the feasibility of the SARR-LOC under sufficient calibration of CV algorithm.
Thus, in the experiments, the positions estimated by the CV-based object detection were regarded as the true object positions.

\subsection{Reference: RSSI-Based Triangulation Localization}
We compared the proposed SARR-LOC with an RSSI-based triangulation localization, which requires three or more RF anchors that measure the received power of the TD.
The received power value at the RF anchor is used to estimate the absolute distance between the TD and at least three RF anchors using path-loss propagation~\cite{kumar2009distance} as follows:
\begin{align}
	\label{equ:simple_pass_loss}
	r = -10 n \log_{10} d +A,
\end{align}
where $r$ indicates the received power value in dB, $n$ is the path-loss exponent, which varies from two in free space to four in indoor environments~\cite{zafari2019survey}, and
$A$ is the received power value at a reference distance from the receiver.
Then triangulation is used to obtain TD's location relative to the RF anchors.

It should be noted that the triangulation method explicitly differ from the proposed SARR-LOC, in terms of the system requirement; the triangulation method requires RSSIs from three or more RF anchors or antenna elements, while the SARR-LOC uses only single-antenna RSSI.
Thus, in this paper, the triangulation method is used as an reference method, whose accuracy is compared to that of SARR-LOC in order-level, and we avoid the detail comparison of the accuracy between the triangulation and the SARR-LOC. 
Moreover, as shown in Section~\ref{ssec:contribution}, due to the deficiency of the single-antenna RSSI in terms of the localization, any RF localization method using only single-antenna RSSI has been proposed.
Thus, we focus on the demonstration of the feasibility of an RF localization method with a single-antenna RSSI.

\subsection{Results}
\subsubsection{Accuracy of Blockage Detection}
This section shows the accuracy of the blockage detection, which estimates the time section when the obstacle overlaps to the FFZ, in the experimental evaluation.
The detection accuracy is evaluated using the estimated blockage sections $\mathcal{T}^\mathrm{sec}$, in which obstacle is estimated to be overlapped to the FFZ,  and the ground-truth blockage sections $\mathcal{T}^\mathrm{gt}$, in which the obstacle actually overlaps to the FFZ.
The ground-truth blockage section are obtained from the ground-truth FFZ derived from the actual measured positions of the TD and RF anchor and the historical obstacle regions.
The ground-truth blockage sections are defined as:
\begin{align}
	\mathcal{T}^\mathrm{gt} = \{t\mid \mathcal{O}(t) \cap \mathcal{F} \neq \emptyset\},
\end{align}
where $\mathcal{F}$ indicates a point set on the ground-truth FFZ.
In this evaluation, the true positive (TP), true negative (TN), false positive (FP), and false negative (FN) are defined as follows:
The TP is the one which estimates that the obstacle overlaps to the FFZ when the obstacle overlaps to the ground-truth FFZ.
The TN is the one which estimates that the obstacle does not overlap to the FFZ when the obstacle does not overlap to the ground-truth FFZ.
The FP is the one which estimates that the obstacle overlaps to the FFZ when the obstacle does not overlap to the ground-truth FFZ.
The FN is the one which estimates that the obstacle does not overlaps to the FFZ when the obstacle overlaps to the ground-truth FFZ.
Given the estimated blockage section $\mathcal{T}^\mathrm{sec}$, we denote the time sections for TP, TN, FP, and FN as 
\begin{align}
	\mathcal{T}^\mathrm{TP} = \{t\mid t \in \mathcal{T}^\mathrm{sec} \cap t \in \mathcal{T}^\mathrm{gt}  \},       \\
	\mathcal{T}^\mathrm{TN} = \{t\mid t \notin \mathcal{T}^\mathrm{sec} \cap t \notin \mathcal{T}^\mathrm{gt}  \}, \\
	\mathcal{T}^\mathrm{FP} = \{t\mid t \in \mathcal{T}^\mathrm{sec} \cap t \notin \mathcal{T}^\mathrm{gt}  \},    \\
	\mathcal{T}^\mathrm{FN} = \{t\mid t \notin \mathcal{T}^\mathrm{sec} \cap t \in \mathcal{T}^\mathrm{gt}  \}.
\end{align}
The length of time section is represented as 
\begin{align}
	\mathrm{len}(\mathcal{T}) = \int_{0}^{t^\mathrm{obs}} \mathbbm{1}(\mathcal{T})\,\mathrm{d}t.
\end{align}

\begin{figure}
	\centering
	\includegraphics[width=0.4\textwidth]{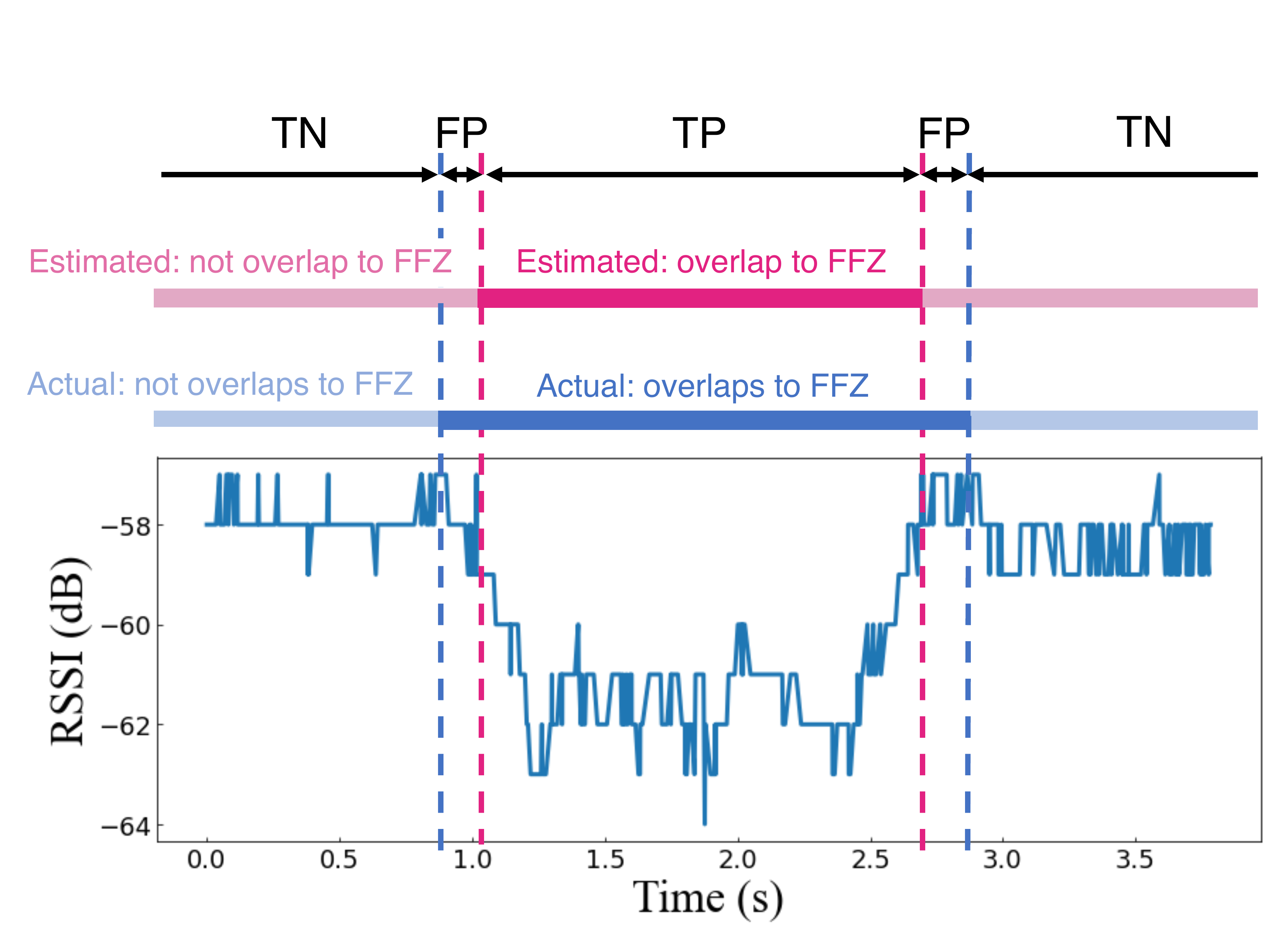}
	\caption{Example of blockage detection result.
	The deep pink band indicates the estimated blockage time section (i.e., the system estimates that obstacle overlaps to the first Fresnel zone).
	The time section, when the RSSI drops, is estimated as a blockage section.
	The deep black band indicates the ground-truth blockage time section (i.e., the obstacle is actually overlaps to the first Fresnel zone).
	}
	\label{fig:blockage_detection_result}
\end{figure}

\begin{table}[t]
	\caption{Ratio of actual and estimated overlaps duration to FFZ to total observation time of 713\,s.}
	\centering
	\scalebox{0.90}[0.90] {
		\begin{tabular}{ccccc}
			\toprule
			                          &        & \multicolumn{2}{c}{Obstacle actually overlap to FFZ}            \\
			                          &        & Yes                                                  & No       \\
			\midrule
			Obstacle is estimated to  & Yes    & $13.9\%$                                             & $2.7\%$  \\
			overlap to FFZ            & No     & $6.0\%$                                              & $76.7\%$ \\
			\midrule
			\multicolumn{2}{c}{Total} & 20.6\% & 79.4\%                                                          \\
			\bottomrule
		\end{tabular}
	}
	\label{tab:miss_detection}
\end{table}

Fig.~\ref{fig:blockage_detection_result} shows an example of the blockage detection results, in the experimental evaluation.
The estimated blockages section (i.e., the time section when the system estimates that obstacle overlaps to the first Fresnel zone) and the ground-truth blockage time section (i.e., the obstacle is actually overlaps to the first Fresnel zone)
are well matched.
The time length of TP, TN, FP, and FN sections were 99\,s, 547\,s, 19\,s, and 43\,s respectively.
The Table~\ref{tab:miss_detection} shows the ratio of time length to the total observation time, respectively, 713\,s.
As shown in Table~\ref{tab:miss_detection}, the system accurately estimates whether the blockage occurred, (i.e., whether the obstacle overlaps to the FFZ or not), for more than 90\% of the entire observation time.
We can conclude that the examined blockage section estimation methods can estimate the blockage section with more than 90\% accuracy.
Thus, we examine the localization accuracy using the blockage section estimation in the following experiments.

\subsubsection{Localization Accuracy Comparison}
\begin{figure}[t]
	\centering
	\includegraphics[width=0.4\textwidth]{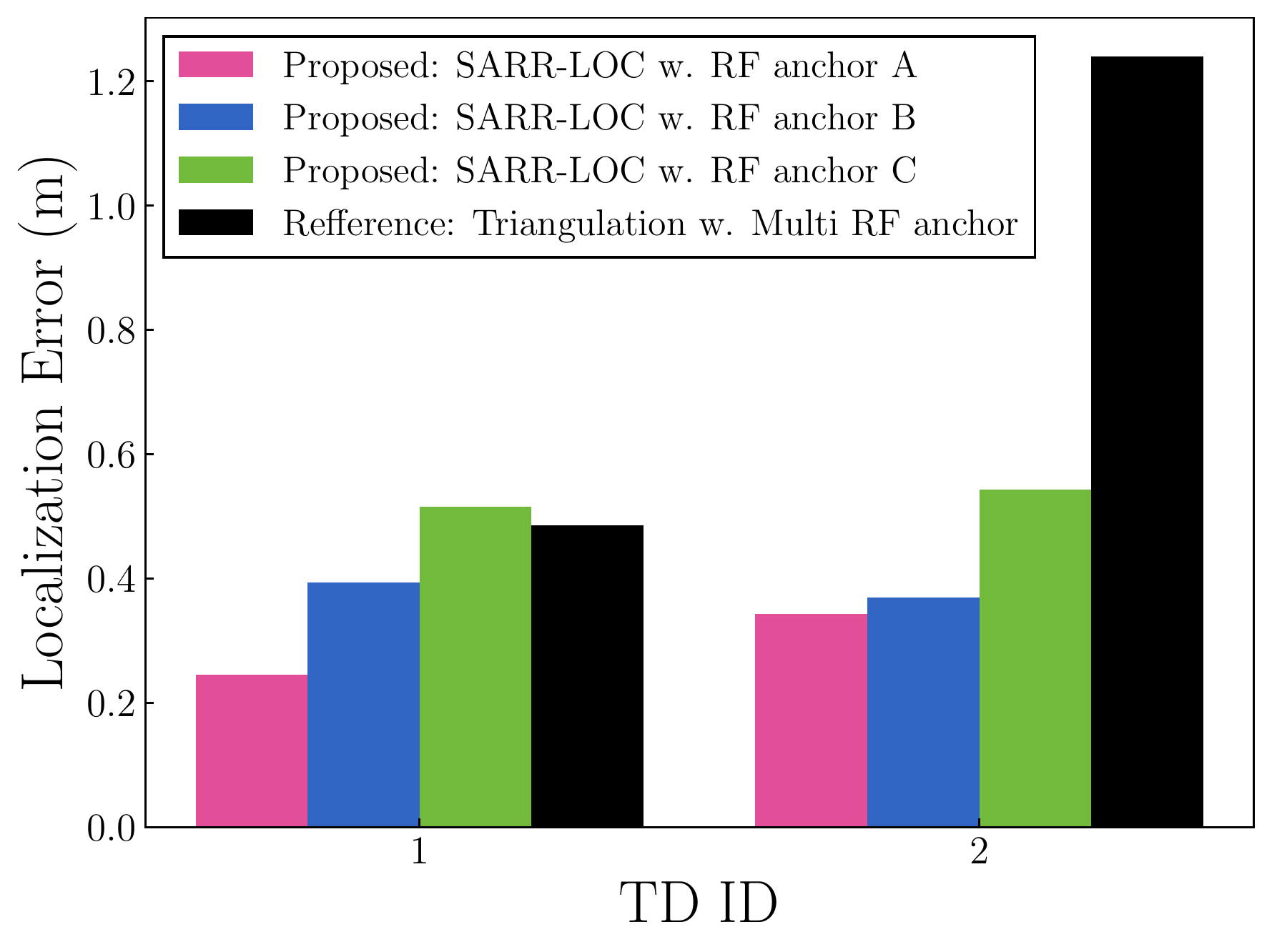}
	\caption{Localization error of proposed SARR-LOC and reference triangulation method.
		The SARR-LOC uses a single RF anchor's RSSI, whereas the triangulation method uses three RF anchor RSSIs.
		In the triangulation method, the path-loss exponent parameter $n$ is set to 2, where the triangulation method achieves a higher performance than the other parameter values.
	}
	\label{fig:main_results}
\end{figure}

\begin{figure*}[t]
	\centering
	\includegraphics[width=0.75\textwidth, page = 3]{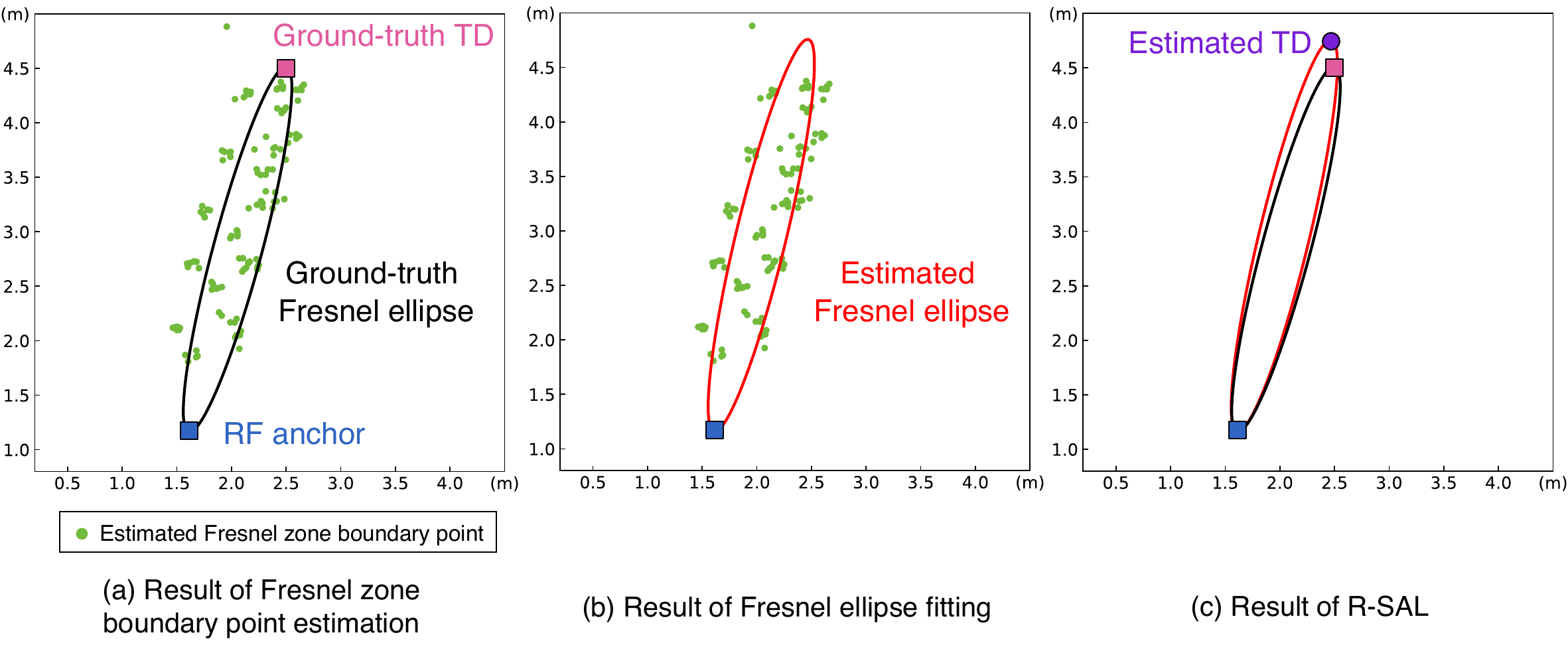}
	\caption{Experimental results of the first Fresnel zone boundary points estimation and first Fresnel ellipse fitting, in case of TD 1 and RF anchor A.
		Fig. (a) depicts the first Fresnel boundary points, which were estimated from the time series of the received power and historical obstacle region.
		Fig. (b) shows the first Fresnel ellipse, which is fitted to the first Fresnel zone boundary points depicted in Fig. (a).
		Moreover, Fig. (c) shows the estimated TD position via the first Fresnel ellipse, which is estimated in Fig. (b), and the ground-truth TD position.
	}
	\label{fig:fresnel_point_estimation}
\end{figure*}

Fig.~\ref{fig:main_results} shows the localization error of the proposed SARR-LOC and the reference triangulation method, where the error is defined as the Euclidian distance between the estimated TD position and the actual measured TD position. 
In the triangulation method, three received powers measured at the RF anchors A, B, and C were used simultaneously.
On the other hand, the SARR-LOC used only the received power time series measured at one of the RF anchors A, B, and C. 
The results show that the proposed SARR-LOC achieves localization errors less than 1.0\,m, which is comparable to that achieved by the triangulation method.
In Fig.~\ref{fig:main_results}, the proposed SARR-LOC achieved a localization error lower than 1.0\,m, whereas the error of the triangulation method is 0.5\,m and 1.2\,m for TD~1 and TD~2, respectively.

Comparing SARR-LOC with the triangulation method, which requires multiple RF anchors, 
in the case of TD~1, the SARR-LOC with RF anchor~A or B achieved a lower error than that of the triangulation method,
whereas the error of SARR-LOC with the RF anchor~C is 17\,cm higher than that of the triangulation method.
The localization error of SARR-LOC with RF anchor C was higher than that with other RF anchors.
This is because of the error of the FFZ boundary point estimation with the RF anchor~C, which is validated in the following section. 
In the case of TD~2, the SARR-LOC achieved a lower error than that of the triangulation method irrespective of the RF anchors.
Thus, although the SARR-LOC does not leverage CSI or RSSIs from multiple anchors, the SARR-LOC  achieves comparable or higher localization accuracy than the conventional triangulation method which requires  multiple anchors.
Therefore, we can conclude that these results demonstrate the feasibility of RSSI localization with single RF anchor with single antenna element, which validates our contribution.

\subsubsection{Result of the first Fresnel Zone Boundary Points Estimation}
More detailed evaluation results are shown in the following sections.
Fig.~\ref{fig:fresnel_point_estimation} shows the results of the FFZ boundary points estimation and first Fresnel ellipse fitting, in the case of TD 1 and the RF anchor A.
Fig.~\ref{fig:fresnel_point_estimation}~(a) depicts the FFZ boundary points estimated from the time series of the received power and the historical obstacle position.
In Fig.~\ref{fig:fresnel_point_estimation}~(a), the estimated FFZ boundary points exist around the ground-truth Fresnel ellipse,
which implies the FFZ boundary points estimation well performs in this setting.
Using the FFZ boundary points shown in Fig.~\ref{fig:fresnel_point_estimation}~(a), the first Fresnel ellipse is estimated, as shown in Fig.~\ref{fig:fresnel_point_estimation}~(b).
Moreover, Fig.~\ref{fig:fresnel_point_estimation}~(c) shows the estimated TD position using the estimated first Fresnel ellipse in Fig.~\ref{fig:fresnel_point_estimation}~(b).
The estimated TD position and first Fresnel ellipse are well fitted to the ground-truth and the localization error is less than 1.0\,m, as shown in Fig.~\ref{fig:main_results}. 

\begin{figure}[t]
	\centering
	\subfloat{\includegraphics[width=0.44\textwidth]{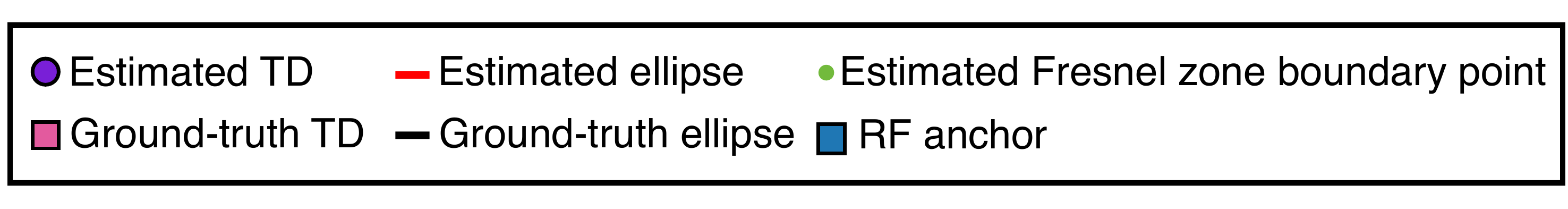}}\\
	\subfloat[(a) RF anchor A]{\includegraphics[width=0.15\textwidth]{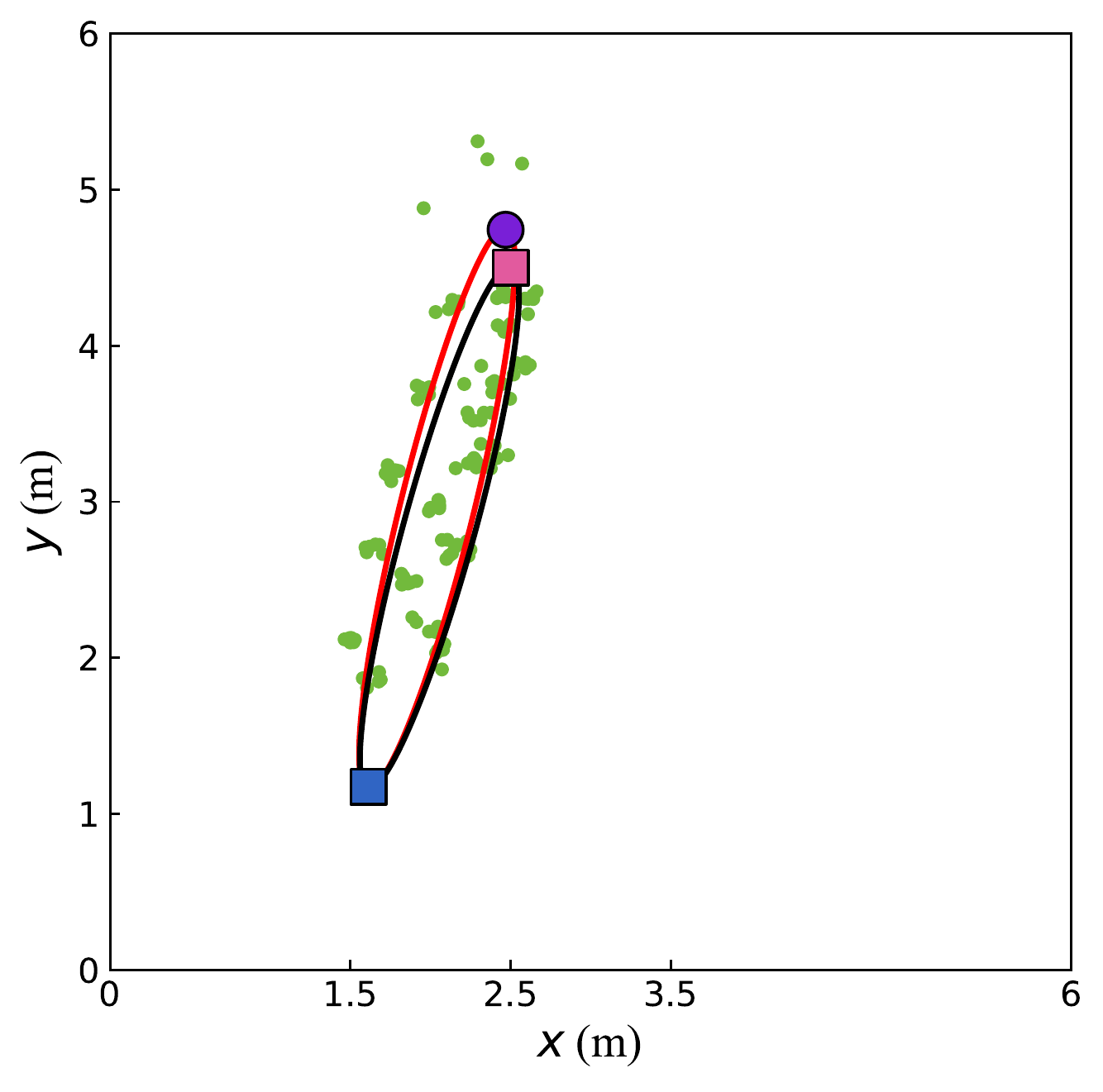}}
	\subfloat[(b) RF anchor B]{\includegraphics[width=0.15\textwidth]{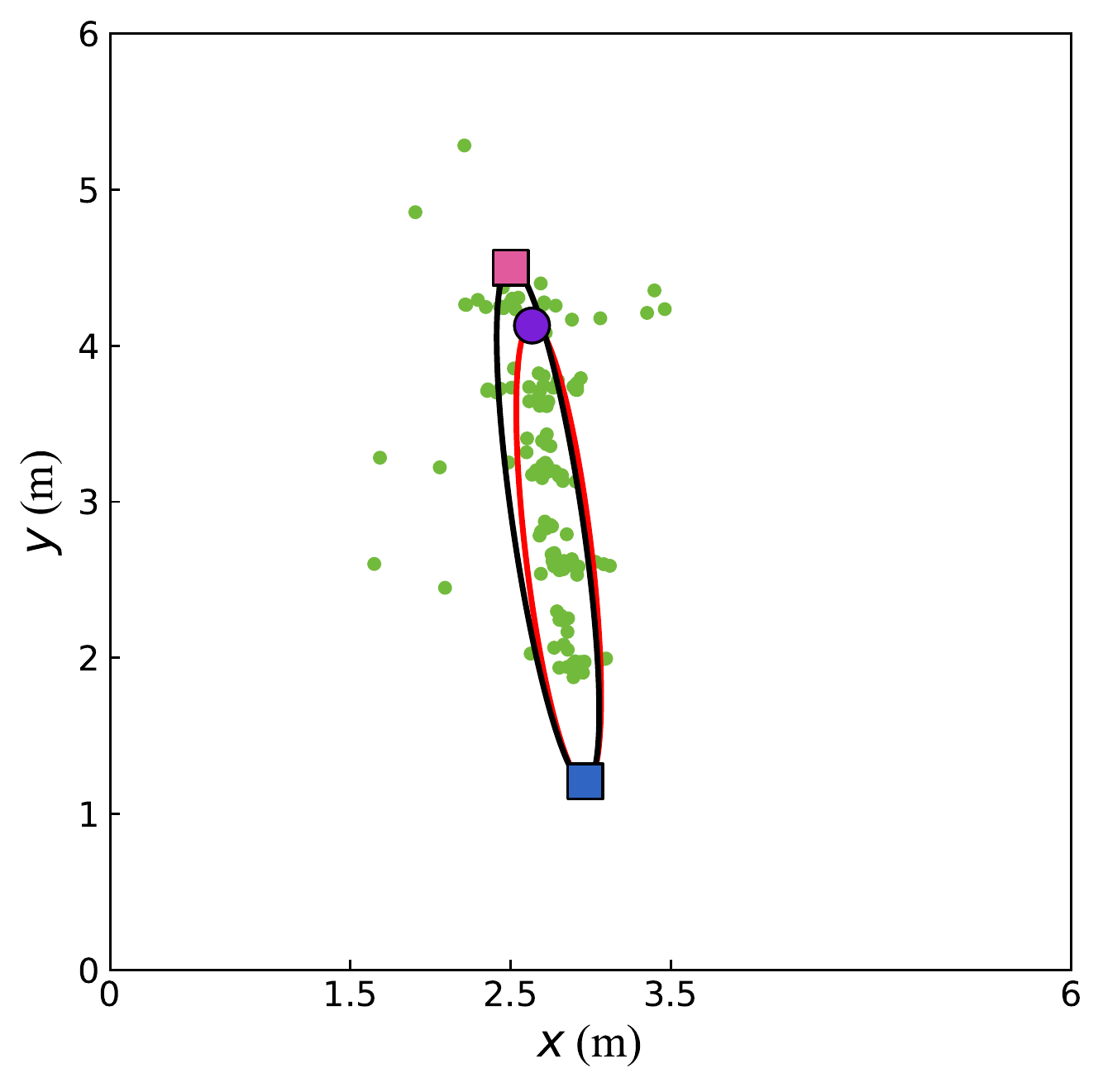}}
	\subfloat[(c) RF anchor C]{\includegraphics[width=0.15\textwidth]{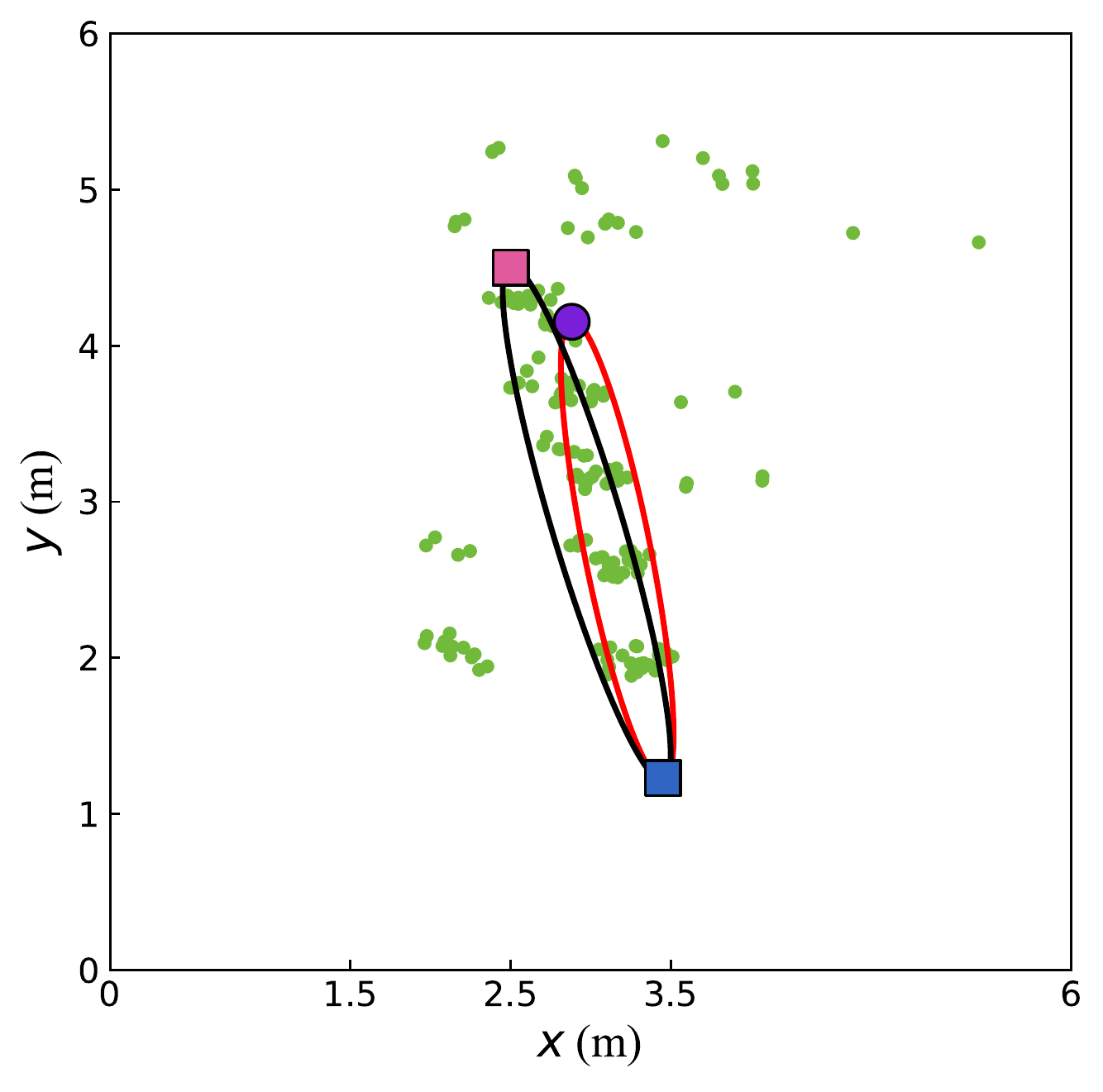}}
	\caption{Experimental results of Fresnel ellipse fitting for TD 1 and each RF anchor.
	}
	\label{fig:fresnel_result}
\end{figure}

Fig.~\ref{fig:fresnel_result} depicts the result of the FFZ boundary points estimation and Fresnel fitting for RF anchor A, B, and C, respectively, in case of TD~1.
Among the three RF anchors, the estimated FFZ boundary points were more scattered from the ground-truth first Fresnel ellipse in the case of RF anchor C.
This is the main reason why in Fig.~\ref{fig:main_results}, the localization error of RF anchor C was higher than that of the other RF anchors.

\begin{figure}[t]
	\centering
	\includegraphics[width=0.3\textwidth]{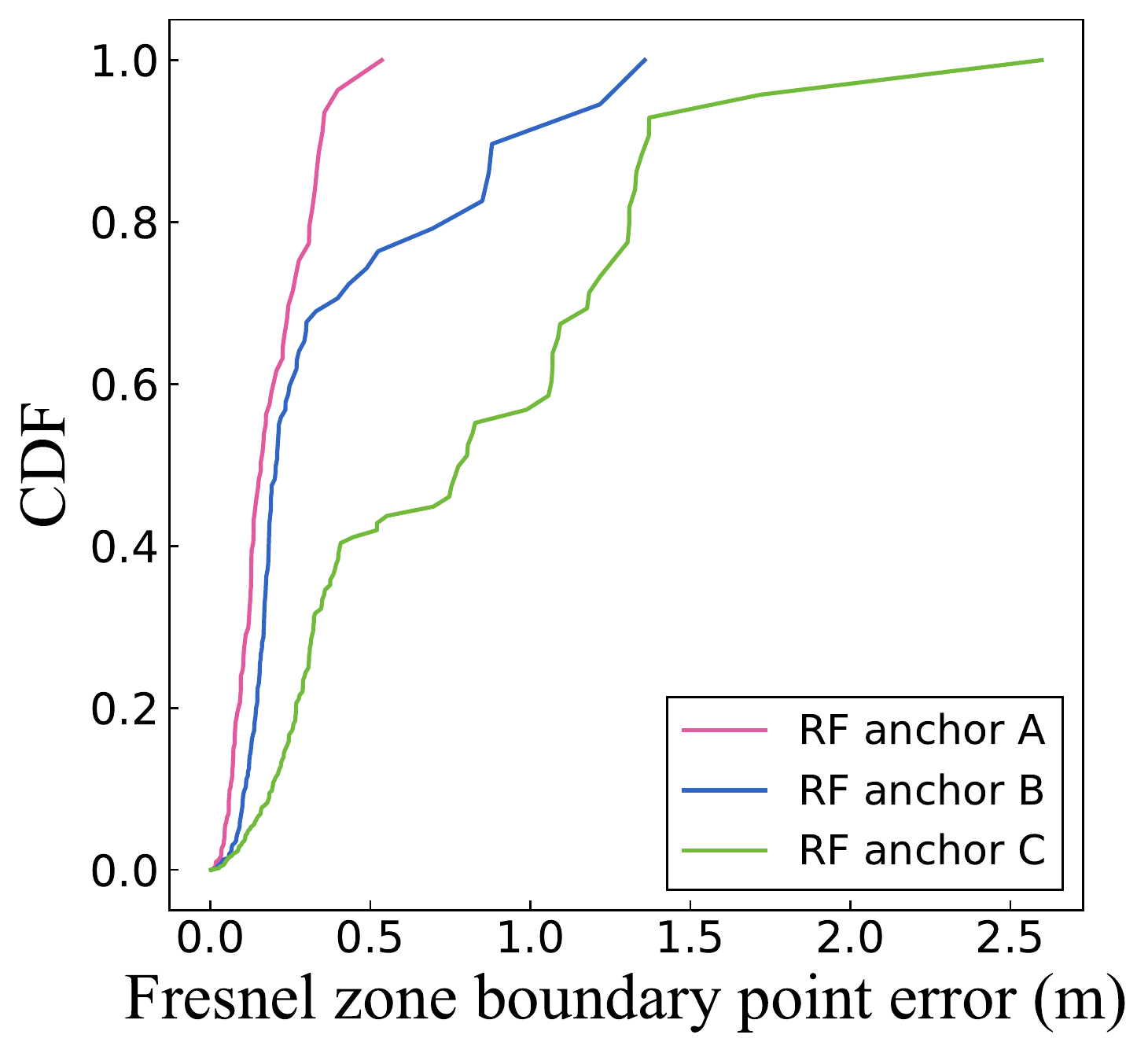}
	\caption{Cumulative distribution function of Euclidian distance from first Fresnel zone boundary points to nearest point on the ground-truth first Fresnel ellipse,
		with TD 1 in case of RF anchors A, B, and C, respectively.}
	\label{fig:Frensel_edge_CDF}
\end{figure}

To evaluate the quantitative error of the FFZ boundary point estimation, we show the cumulative distribution function (CDF) of the distance of each FFZ boundary points to the ground-truth Fresnel ellipse in Fig.~\ref{fig:Frensel_edge_CDF}.
The distance is calculated as the Euclidian distance from the FFZ boundary points to the nearest point on the ground-truth first Fresnel ellipse.
The longer the distance, the larger the error of the FFZ boundary points. 
Consistent with the visualization shown in Fig.~\ref{fig:fresnel_result}, the distance of RF anchor C was higher than that of the other RF anchors.
This implies that the FFZ boundary points of RF anchor C are more useless than the other RF anchors, which results in the poorest localization accuracy.

\subsubsection{Effect of Object Moving Line and Observation Duration}
\begin{figure}[t!]
	\centering
	\includegraphics[width=0.285\textwidth]{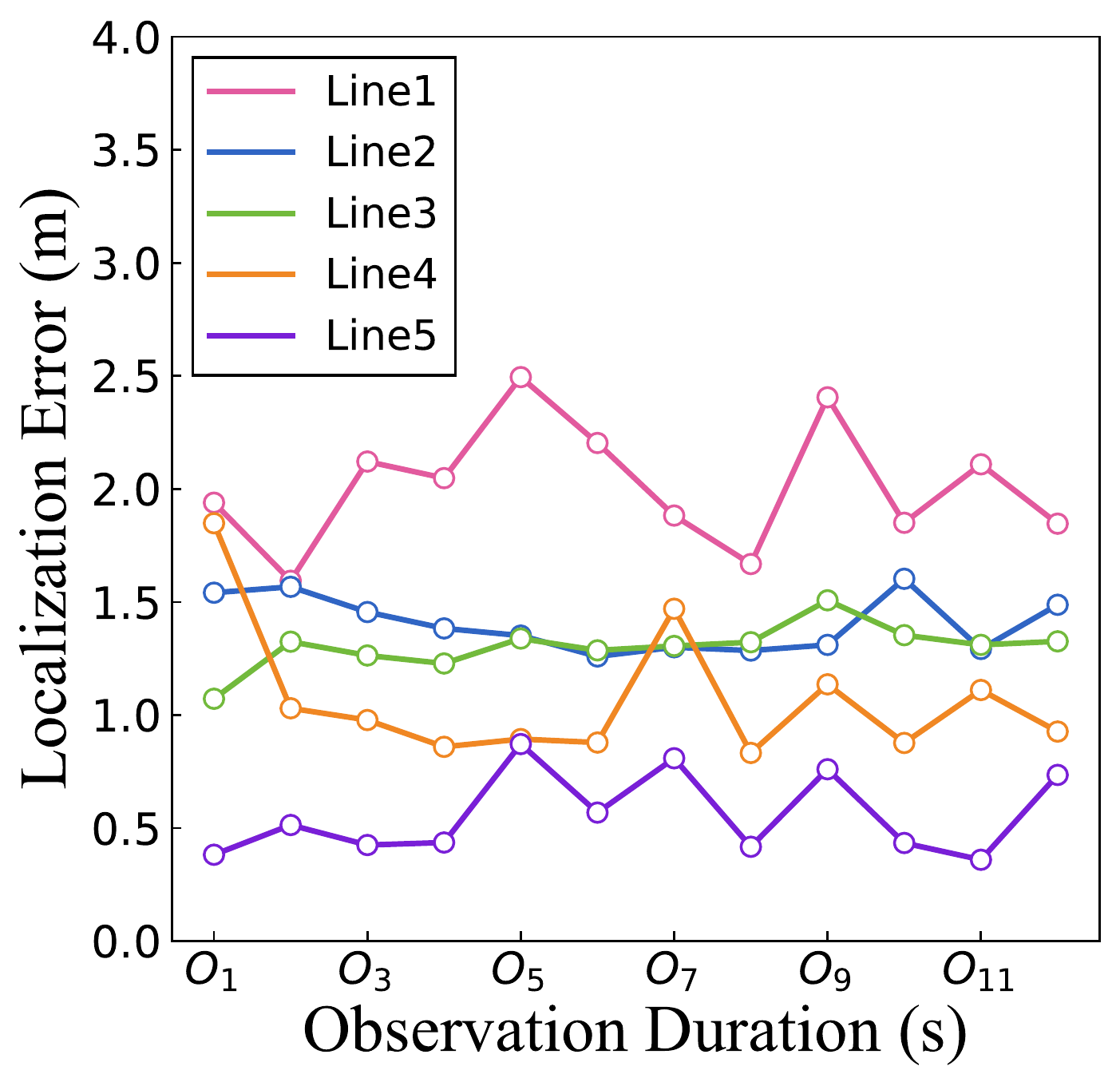}
	\caption{
		Average localization error of SARR-LOC for each obstacle moving line.
		The observation duration $o_b$ indicates the time duration in which the blockage occurs $b$ times in each moving line.
	}
	\label{fig:path_error_line}
\end{figure}

Fig.~\ref{fig:path_error_line} shows the average localization error of SARR-LOC for each obstacle moving line, and for the number of detected blockage events, respectively.
The average localization error was calculated by averaging the six localization errors, that is, the six localization errors for the combination of three RF anchors and two TDs.
We define an observation duration $o_b$ where the blockage occurs $b$ times in each moving line and $o_{b-1} < o_b$.
Although the number of blockage events which actually occurred  was $b$ in $o_b$ duration in the experiment, the number of detected blockages for each pair of a TD and an RF anchor could be less or more than $b$ because of  FP and FN.
Therefore, the localization was conducted with the dataset obtained for each moving line and pair of the TD and RF anchors in $o_b$ duration, and then the localization error was calculated.
Note that we could not obtain any localization result for line 6,
because line 6 does not intersect with the FFZ, which results in no blockage events.

Comparing the error between the obstacle moving lines, as the obstacle blocks the FFZ closer to the TD, the localization error is improved.
Moreover, the error of line 1 is the highest among the five lines irrespective of the observation duration.
This is because, as a line close to the RF anchor, the first Fresnel ellipse estimation is more sensitive to the noise of the FFZ boundary points,
which is introduced by the CV object detection and blockage time section detection. 
Because the FFZ boundary points are generated on the obstacle moving lines and the first Fresnel ellipse is estimated using the position of the RF anchor and the FFZ boundary points, as the lines close to the RF anchor,
the points used for first Fresnel ellipse fitting exist on a smaller area.
In general, owing to the small fraction of the data points in the domain of the original function, the accuracy of function fitting is more sensitive to  noise.
Thus, as the line goes closer to the RF anchor, the first Fresnel ellipse fitting can use the data point in a smaller area, which increases the sensitivity of the first Fresnel ellipse fitting to the noises.

Comparing $o_b$, the localization using a single blockage achieved comparable accuracy to that obtained in $o_{12}$, where the blockage occurred 12 times. 
This implies that a few FFZ boundary points can represent the entire shape of the FFZ. 
It can be concluded that the proposed method provides accurate localization using a few blockage events, which can be completed within a few seconds.

\section{Conclusion}
\label{sec:conclusion}
We showed the feasibility of the single-antenna and single-anchor RSSI based localization, termed SARR-LOC.
The fundamental idea of the proposed SARR-LOC was to leverage a CV technique to compensate for deficient RF information.
More specifically, by combining the CV object detection with a time series of RSSI, we recorded blockage points, which were referred to as the positions of objects when a FFZ blockage occurred. 
These historical blockage points reflect the position and size of the FFZ, and thus, the device can be localized.
The experimental evaluation revealed that the proposed SARR-LOC achieved decimeter-level localization accuracy in an indoor environment, which was comparable to or higher than that in a previous triangulation with three RF anchors' RSSIs.

Future work will include adapting the proposed localization method under multiple moving obstacles, where we should identify which obstacles cause blockage events.
Moreover, an orientation for future work is to develop a single-antenna RSSI-based localization method applicable to a case where a line-of-sight path is unavailable.

\section{Appendix}
\subsection{Empirical Improvement of First Fresnel Zone Fitting}
\label{ssec:deatil:fitting}
\begin{figure}[t!]
	\centering
	\includegraphics[width=0.47\textwidth]{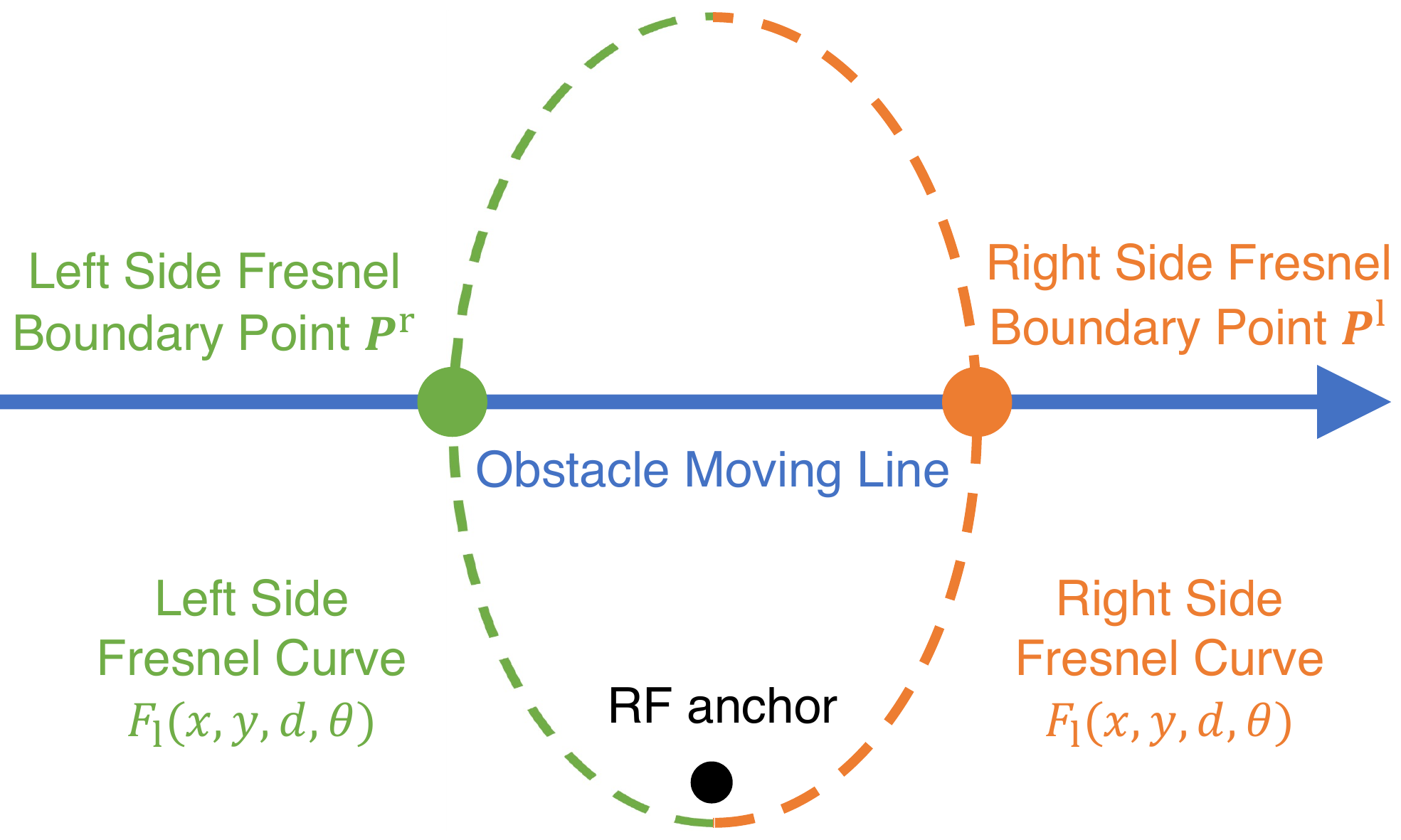}
	\caption{
		First Fresnel zone boundary points separation method. 
		The right-and left-side first Fresnel zone boundary points are used for fitting the right-side first Fresnel curve and left-side first Fresnel curve, respectively.
	}
	\label{fig:fresnel_split_2}
\end{figure}

To improve the accuracy of SARR-LOC, we used two FFZ boundary points, which were generated at one blockage event differently.
In a blockage event, two FFZ boundary points $f^\mathrm{s}(\mathcal{O}(t^\mathrm{s}))$ and $f^\mathrm{e}(\mathcal{O}(t^\mathrm{e}))$ are generated.
As shown in Fig.~\ref{fig:fresnel_split_2}, assuming the obstacle moving line crosses a line that passes through the TD and RF anchor, one of the two points exists on the right side of the line that passes through the TD and RF anchors.
Conversely, the other point exists on the left side.
We identify whether the side of the point exists, and then leverage the information and use in different ways. 
In more detail, we divide the first Fresnel ellipse equation into two curves, and the point on the right side is used for fitting a curve, and the other is used for the other curve.

Let us denote the two subsets of the FFZ boundary points divided by the center line as the left subset $\bm{P}^\mathrm{l}$ and right subset $\bm{P}^\mathrm{r}$.
The first Fresnel ellipse equation is divided into the two curves as follows:
\begin{align}
	F_{\mathrm{r}}(x,y,d,\theta) & = y'+ \frac{\sqrt{\lambda(4d+\lambda)}}{4}\sqrt{1-\frac{(4x'-2d)^2}{(2d + \lambda)^2}}=0, \label{equ:right_curve} \\
	F_{\mathrm{l}}(x,y,d,\theta) & = y' -\frac{\sqrt{\lambda(4d+\lambda)}}{4}\sqrt{1-\frac{(4x'-2d)^2}{(2d + \lambda)^2}}=0,\label{equ:left_curve}   \\
	x'                           & \coloneqq x \cos \theta + y \sin \theta                                                                           \\
	y'                           & \coloneqq y \cos \theta - x \sin \theta
\end{align}
where \eqref{equ:right_curve} and \eqref{equ:left_curve} indicate the right-and left-side curves, respectively.
Thus, the optimization problem (22) in the main paper is divided into two function optimizations as follows:
\begin{align}
	\label{equ:easy_optimization}
	d^{*}, \theta^{*} =\min_{d, \theta}
	\frac{1}{|\bm{P}^\mathrm{r}|} \sum_{(x, y) \in \bm{P}^\mathrm{r}} \{F_{\mathrm{r}}(x,y,d,\theta)\}^2 \notag \\
	+\frac{1}{|\bm{P}^\mathrm{l}|} \sum_{(x, y) \in \bm{P}^\mathrm{l}} \{F_{\mathrm{l}}(x,y,d,\theta)\}^2.
\end{align}

To divide the FFZ boundary points into two subsets, the blockage event starts and end points are associated.
Assuming the blockage event starts at $\bm{p}^\mathrm{s}$ and ends at $\bm{p}^\mathrm{e}$, one is categorized into the right subset and another to the left subset based on the relative position of the center line. 
Using $\bm{p}^\mathrm{s}$ and $\bm{p}^\mathrm{e}$, the center line is estimated as follows;
\begin{align}
	\label{equ:bisection}
	y = g(x) =\frac{p^\mathrm{b}_{\mathrm{y}}}{p^\mathrm{b}_{\mathrm{x}}}x,
\end{align}
where $\bm{p}^{\mathrm{b}}$ denotes a bisection point of $\bm{p}^\mathrm{s}$ and $\bm{p}^\mathrm{e}$.
Based on the relative positions of the estimated center line, $\bm{p}^\mathrm{s}$ and $\bm{p}^\mathrm{e}$ are categorized as follows:
\begin{align}
	\label{equ:separate_rightleft}
	\begin{cases}
		\bm{p}^\mathrm{s} \in \bm{P}^\mathrm{l},  \bm{p}^\mathrm{e} \in \bm{P}^\mathrm{r} \  & \text{if}\ g(p^\mathrm{e}_\mathrm{x}) > p^\mathrm{e}_\mathrm{y}; \\
		\bm{p}^\mathrm{s} \in \bm{P}^\mathrm{r},  \bm{p}^\mathrm{e} \in \bm{P}^\mathrm{l}\   & \text{otherwise}.
	\end{cases}
\end{align}
Note that this separation is performed prior to the obstacle size compensation.

\subsection{Mathematical Explanation of First Fresnel Zone Boundary Point Estimation from Blockage Starts and Ends Times}
This section explain the mathematical explanation for the FFZ boundary point estimation from obstacle region at the blockage starts or ends times.
For instance, the FFZ boundary point at the time blockage starts is estimated as $\bm{e}^\mathrm{+}$ and that at  the time blockage ends is estimated as $\bm{e}^\mathrm{-}$, respectively, due to the reason detailed as following paragraphs.

First, we consider the FFZ boundary point at the time blockage starts.
When the blockage starts, the obstacle enters to the FFZ.
Formally, only one point in the obstacle region overlaps with the FFZ, where the point exists on the FFZ boundary. 
Given the blockage starts time $t^\mathrm{s}$, infinitesimal value $\delta t$, and ground-truth FFZ $\mathcal{F}$, the obstacle region satisfies the following conditions:
\begin{align}
	\label{equ:A2_case}
	\begin{cases}
		\mathcal{O}(t^\mathrm{s} - \delta t) \cap \mathcal{F}  = \emptyset, \\
		\mathcal{O}(t^\mathrm{s}) \cap \mathcal{F} = \{\bm{p}^\mathrm{s}\},
	\end{cases}
\end{align}
where $\bm{p}^\mathrm{s}$ indicates the FFZ boundary point.

To obtain $\bm{p}^\mathrm{s}$, a set difference of \eqref{equ:A2_case} is calculated as: 
\begin{align}
	\{\bm{p}^\mathrm{s}\} \setminus \emptyset & = (\mathcal{O}(t^\mathrm{s}) \cap \mathcal{F}) \setminus(\mathcal{O}(t^\mathrm{s} - \delta t) \cap \mathcal{F}) \\
	\label{equ:A2_con}
	\{\bm{p}^\mathrm{s}\}                     & =(\mathcal{O}(t^\mathrm{s}) \setminus \mathcal{O}(t^\mathrm{s} - \delta t)) \cap \mathcal{F},
\end{align}
A necessary condition to satisfy \eqref{equ:A2_con} is that $\bm{p}^\mathrm{s}$ satisfies an condition:
\begin{align}
	\label{equ:A2_con2}
	\bm{p}^\mathrm{s} \in \mathcal{O}(t^\mathrm{s}) \setminus \mathcal{O}(t^\mathrm{s} - \delta t).
\end{align}
Considering $\delta t$ is infinitesimal, $\bm{n}(t)$ is consist from $t - \delta t$ to $t$, then we obtain 
\begin{align}
	\mathcal{O}(t^\mathrm{s} - \delta t) & =\left\{\bm{p}^\mathrm{cnt}(t-\delta t) + k\bm{n}(t-\delta t) \relmiddle| |k| < \frac{w}{2}\right\}       \\
	                                     & =  \left\{\bm{p}^\mathrm{cnt}(t) -\delta t v\bm{n}(t) + k\bm{n}(t) \relmiddle| |k| < \frac{w}{2}\right\},
\end{align}
where $v$ indicate velocity of the obstacle.
As explained in the main paper, 
\begin{align}
	\mathcal{O}(t^\mathrm{s})=\left\{\bm{p}^\mathrm{cnt}(t) + k\bm{n}(t) \relmiddle|   |k| < \frac{w}{2}\right\},
\end{align}
then the set difference is denoted as 
\footnotesize
\begin{align}
	\mathcal{O}(t^\mathrm{s}) \setminus \mathcal{O}(t^\mathrm{s} - \delta t) & = \left\{\bm{p}^\mathrm{cnt}(t) -\tau v\bm{n}(t) + \frac{w}{2}\bm{n}(t) \relmiddle| 0 \leq \tau < \delta t \right\} \\
	                                                                         & =\left\{ \bm{p}^\mathrm{cnt}(t) + \frac{w}{2}\bm{n}(t) \right\}                                                     \\
	\label{equ:mou_namae_kanngaeruno_mendoi}
	                                                                         & =\{\bm{e}^\mathrm{+}\}.
\end{align}
\normalsize
Using \eqref{equ:mou_namae_kanngaeruno_mendoi}, the condition \eqref{equ:A2_con2} is represented as $\bm{p}^\mathrm{s} \in \{\bm{e}^\mathrm{+}\}$, 
and thereby we obtain $\bm{p}^\mathrm{s} = \bm{e}^\mathrm{+}$.

In case of the FFZ boundary point at the time blockage ends, the obstacle leaves from the FFZ.
Thus, the condition \eqref{equ:A2_case} is converted to as follows:
\begin{align}
	\begin{cases}
		\mathcal{O}(t^\mathrm{e} - \delta t) \cap \mathcal{F}  = \{\bm{p}^\mathrm{e}\}, \\
		\mathcal{O}(t^\mathrm{e}) \cap \mathcal{F} = \emptyset,
	\end{cases}
\end{align}
where $\bm{p}^\mathrm{e}$ and $t^\mathrm{e}$ indicate the FFZ boundary point and the blockage ends time, respectively.
In the same way to the discussion above, we obtain $\bm{p}^\mathrm{e} = \bm{e}^\mathrm{-}$.

\bibliographystyle{IEEEtran}
\bibliography{main_1.bbl}

\begin{IEEEbiography}
	[{\includegraphics[width=1in, height=1.25in, clip, keepaspectratio]{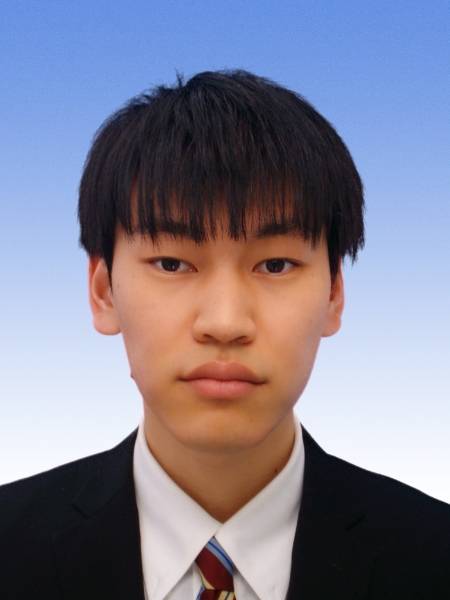}}]
	{Tomoya~Sunami}
	received the B.E. degree in electrical and electronic engineering from Kyoto University in 2020.
	He is currently studying toward the M.I. degree at the Graduate School of Informatics, Kyoto University.
	He is a student member of the IEEE.
\end{IEEEbiography}

\begin{IEEEbiography}
	[{\includegraphics[width=1in, height=1.25in, clip, keepaspectratio]{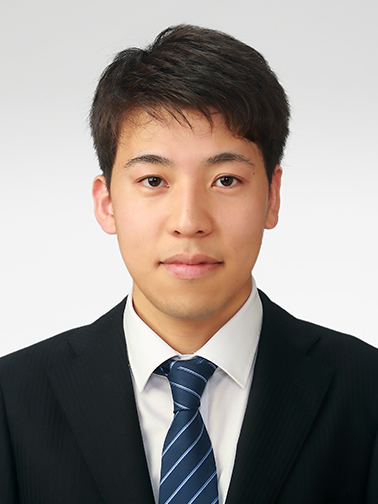}}]
	{Sohei~Itahara}
	received the B.E. degree in electrical and electronic engineering from Kyoto University in 2020.
	He is currently studying toward the M.I. degree at the Graduate School of Informatics, Kyoto University.
	He is a student member of the IEEE.
\end{IEEEbiography}

\begin{IEEEbiography}[{\includegraphics[width=1in, height=1.25in, clip, keepaspectratio]{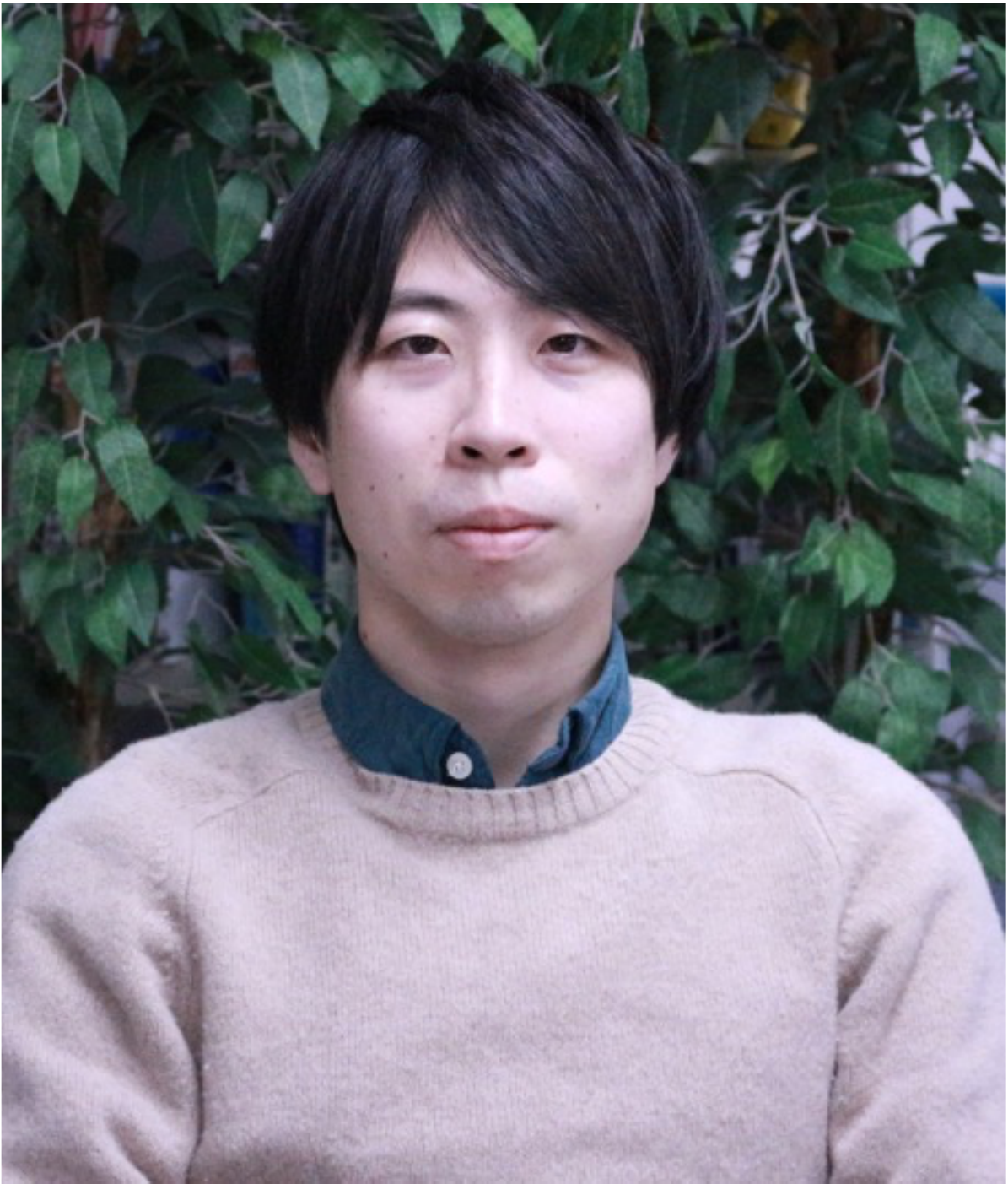}}]{Yusuke~Koda}
	received the B.E. degree in electrical and electronic engineering from Kyoto University in 2016 and the M.E. degree at the Graduate School of Informatics from Kyoto University in 2018.
	In 2019, he visited Centre for Wireless Communications, University of Oulu, Finland to conduct collaborative research.
	He is currently studying toward the Ph.D. degree at the Graduate School of Informatics from Kyoto University.
	He was a Recipient of the Nokia Foundation Centennial Scholarship in 2019.
	He received the VTS Japan Young Researcher's Encouragement Award in 2017.
	He is a member of the IEICE and a member of the IEEE.
\end{IEEEbiography}

\begin{IEEEbiography}
	[{\includegraphics[width=1in, height=1.25in, clip, keepaspectratio]{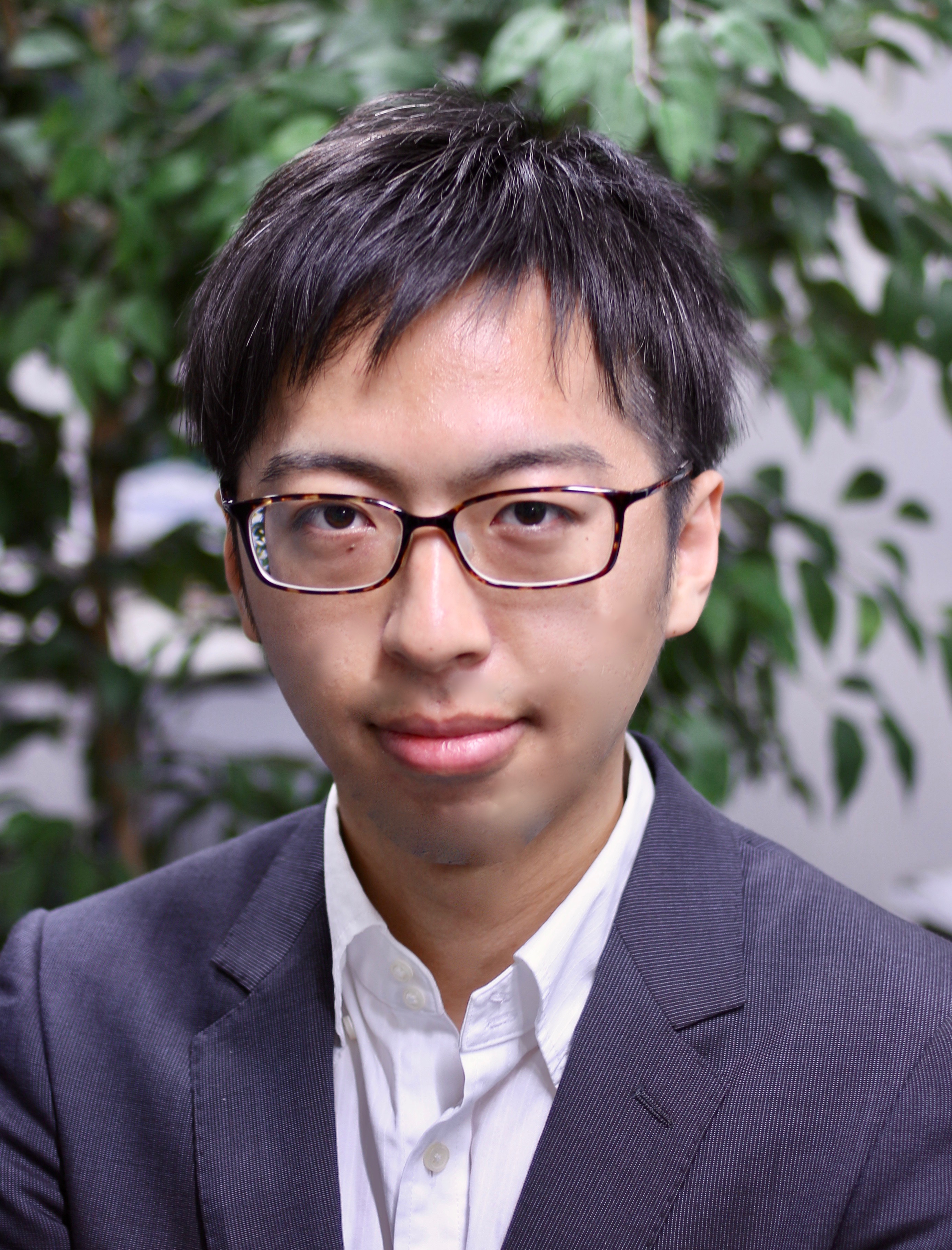}}]
	{Takayuki~Nishio}
	has been an associate professor in the School of Engineering, Tokyo Institute of Technology, Japan, since 2020.
	He received the B.E.\ degree in electrical and electronic engineering and the master's and Ph.D.\ degrees in informatics from Kyoto University in 2010, 2012, and 2013, respectively.
	He had been an assistant professor in the Graduate School of Informatics, Kyoto University from 2013 to 2020.
	From 2016 to 2017, he was a visiting researcher in Wireless Information Network Laboratory (WINLAB), Rutgers University, United States.
	His current research interests include machine learning-based network control, machine learning in wireless networks, and heterogeneous resource management.
\end{IEEEbiography}

\begin{IEEEbiography}
	[{\includegraphics[width=1in, height=1.25in, clip, keepaspectratio]{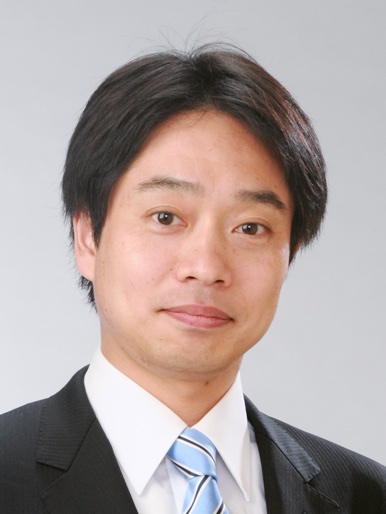}}]
	{Koji~Yamamoto}
	received the B.E. degree in electrical and electronic engineering from Kyoto University in 2002, and the M.E. and Ph.D. degrees in Informatics from Kyoto University in 2004 and 2005, respectively.
	From 2004 to 2005, he was a research fellow of the Japan Society for the Promotion of Science (JSPS).
	Since 2005, he has been with the Graduate School of Informatics, Kyoto University, where he is currently an associate professor.
	From 2008 to 2009, he was a visiting researcher at Wireless@KTH, Royal Institute of Technology (KTH) in Sweden.
	He serves as an editor of IEEE Wireless Communications Letters from 2017 and the Track Co-Chairs of APCC 2017 and CCNC 2018.
	His research interests include radio resource management and applications of game theory.
	He received the PIMRC 2004 Best Student Paper Award in 2004, the Ericsson Young Scientist Award in 2006.
	He also received the Young Researcher's Award, the Paper Award, SUEMATSU-Yasuharu Award from the IEICE of Japan in 2008, 2011, and 2016, respectively, and IEEE Kansai Section GOLD Award in 2012.
\end{IEEEbiography}

\vfill

\end{document}